\documentclass[journal=jacsat,manuscript=article]{achemso}
\usepackage{amsmath}
\usepackage{physics}
\usepackage{float}
\usepackage[utf8]{inputenc}
\usepackage{tablefootnote}
\usepackage{xcolor}
\usepackage{soul}
\usepackage{multirow}
\usepackage{setspace}

\usepackage[most]{tcolorbox}
\newtcolorbox{highlighted}{colback=yellow,coltext=red,breakable}
\usepackage{lipsum}
\usepackage[version=3]{mhchem} 
\usepackage{makecell}
\usepackage{microtype}

\author{Soumyadip Hazra}
\affiliation{CAMRIE, Indian Institute of Science Education and Research Thiruvananthapuram, Kerala 695551, India}
\alsoaffiliation{School of Physics, Indian Institute of Science Education and Research Thiruvananthapuram, Kerala 695551, India}
\email{soumyadip22@iisertvm.ac.in}
\author{Sraboni Dey}
\affiliation{School of Physics, Indian Institute of Science Education and Research Thiruvananthapuram, Kerala 695551, India}
\author{Arijit Kayal}
\affiliation{J. Heyrovsky Institute of Physical Chemistry, Czech Academy of Sciences, Dolejskova 2155/3, 18200 Prague 8, Czech Republic}
\alsoaffiliation{School of Physics, Indian Institute of Science Education and Research Thiruvananthapuram, Kerala 695551, India}
\author{Narendra Shah}
\affiliation{School of Physics, Indian Institute of Science Education and Research Thiruvananthapuram, Kerala 695551, India}
\author{Renjith Nadarajan}
\affiliation{School of Physics, Indian Institute of Science Education and Research Thiruvananthapuram, Kerala 695551, India}
\author{Joy Mitra}
\email{j.mitra@iisertvm.ac.in}
\affiliation{School of Physics, Indian Institute of Science Education and Research Thiruvananthapuram, Kerala 695551, India}
\alsoaffiliation{CAMRIE, Indian Institute of Science Education and Research Thiruvananthapuram, Kerala 695551, India}

\title
{Machine Learning Assisted Reconstruction of Local Electronic Structure of Non-Uniformly Strained \texorpdfstring{MoS$_2$}{MoS2}}
\SectionNumbersOn
\usepackage{microtype}
\usepackage[colorlinks=true,allcolors=blue]{hyperref}
\begin{document}
\begin{spacing}{1}
\begin{abstract}
Wrinkles and nanobubbles are an integral and often unavoidable part of integrating 2D van der Waals semiconductors into actual device architectures. Despite their ubiquitous nature, quantitative correlation between such spatially non-uniform strain and modifications to the local electronic structure remains challenging. Here, density functional theory is combined with a recurrent neural network to reconstruct the local electronic structure of monolayer MoS$_2$ from strain maps derived from atomic force microscopy (AFM) topography and Raman spectral maps. 
The analysis reveals that biaxial bending induced strain is significantly more effective than both uniaxial bending or in-plane strain in modifying electronic and dielectric properties. A $\sim$ 0.35\% strain induced by biaxial bending results in $\sim$ 22\% reduction in band gap and $\sim$ 7\% increase in dielectric constant, compared to a $\sim$ 5\% reduction in band gap and $\sim$ 1\% increase in dielectric constant under comparable uniaxial bending. The modified band structure reveals band edge states that concentrate charge in regions of high curvature or strain. While conductive AFM measurements indicate increased local conductance (carrier density) at wrinkles and nanobubbles, the spatial band gap maps predicted by the model are validated against experimental photoluminescence peak energy maps. The results indicate that strained features like wrinkles and nanobubbles commonly present in real devices influence the band gap, carrier distribution, and dielectric response, which favourably affects electrical transport in such systems. The framework developed here can be readily extended to other 2D materials and heterostructures, offering a computationally efficient route for studying and exploiting strain effects.
\end{abstract}
\section{\label{sec:Intro}
Introduction}
Precise control of carrier density and mobility in semiconductors is central to the development of high-performance, energy-efficient electronic and optoelectronic devices. The limitations of conventional doping strategies used in bulk semiconductors have been mitigated
in two-dimensional electron gas (2DEG) systems, where quantum confinement and spatial separation of the active channel from dopants enhance mobility, along with high gate tunability. Such architectures underpin modern nanoscale transistors and quantum devices capable of high-frequency operation with ultralow power consumption \cite{Radisavljevic2011-lc}. The isolation of graphene \cite{novoselov2004electric} catalysed the emergence of a broader class of van der Waals (vdW) materials in which charge carriers are intrinsically confined to atomic-scale thicknesses. Reduced dielectric screening, strong Coulomb interactions, and symmetry-driven band structures in these systems gave rise to a wealth of emergent quantum phenomena, including strongly bound excitons\cite{Liang2014-lu, Sajjad2018-ur}, valley polarisation\cite{Kayal2022-rq, Ho2020-hk, Gopalan2025-vd, Zhang2022-po, Kumar2025-ii}, excitonic condensation\cite{Gao2023-yl, Moon2025-vx}, and correlated insulating and superconducting phases\cite{Dong2022-qg, Jamwal2026-oz} that have no direct bulk counterparts. Within the 2D materials family, transition metal dichalcogenides (TMDCs) are particularly compelling due to their semiconducting nature, strong in-plane covalent bonding, and weak interlayer van der Waals interactions, which enable both mechanical exfoliation and controlled growth in atomically thin layers\cite{Wang2020-ts, Kang2022-ua}.
Molybdenum disulfide (MoS$_2$), a prototypical TMDC, exhibits polymorphism (1T, 1H, 2H, and 3R) \cite{sam2020probing} and a pronounced thickness-dependent band structure, transforming from an indirect to a direct bandgap semiconductor at the monolayer limit \cite{kuc2011influence,yun2012thickness}. Here, strain engineering has emerged as a powerful controllable tool to modulate the local electronic and phononic structure and the ensuing electrical, optical and opto-electronic properties, crucially without introducing chemical dopants.
Experimental results for uniaxially and biaxially stretched monolayer (ML) MoS$_2$ show that the bandgap energy ($E_g$) decreases $\sim$ 45-70 meV/\% of uniaxial strain \cite{conley2013bandgap, he2013experimental, wang2022first} and by $\sim$ 100 meV/\% for biaxial strain \cite{plechinger2015control, mohammad2013first, yu2015phase, li2015bandgap, lopez2016band}. 
However, actual deformations in 2D flakes immobilised on substrates are more complex, with local biaxial bending of the lattice evidenced by nanobubbles, wrinkles, and other buckled structures that nucleate spatially non-uniform strain across the flakes.
Although the kinetics of formation of these deformations are directed by the specifics of the fabrication process, their equilibrium configuration is primarily governed by a competitive interplay between the material's adhesion with the substrate and its bending modulus. The effect of such non-uniform strain has been explored experimentally using elastomeric substrates\cite{castellanos2013local}, nanoindentation \cite{Huang2011-fd}, patterned substrates\cite{kayal2023mobility, yanev2024programmable, yang2025geometrically, wang2019strain}, etc., resulting in substantial modification to the physical and chemical properties of the flakes. 
These high-strain deformations also lead to local doping, phonon quenching, and modifications to the dielectric environment, which collectively contribute to improved carrier transport \cite{ng2022improving, kayal2023mobility} and enhanced catalytic properties \cite{Nadarajan2024-fl} in ML MOS$_2$.  
Recent developments have also demonstrated strain-tunable TMDC electronics, where controlled deformation enables local band alignment\cite{Ai2025-ok}, exciton funnelling\cite{Moon2020-ss, Qian2025-jw}, piezoresistive switching\cite{Manzeli2015-dr, Nakamura2018-jd}, and dynamically reconfigurable device behavior.
In comparison, first-principles calculation based investigations on changes in electronic properties  in non-uniformly strained 2D TMDC systems, due to local biaxial bending, have received limited attention \cite{yu2016bending, gonzalez2018bending,Gupta2022-mk,Gupta2025TMD}. 

Here, the influence of local biaxial strain on the structural, electronic, and dielectric properties of ML MoS$_2$ is investigated using density functional theory (DFT) calculations\cite{kohn1965self}. The deformations are simulated by bending the atomically thin lattice into a three-dimensional Gaussian feature, mimicking experimentally observed surface topographies. The relaxed atomic lattice is used to estimate local strain, and the electronic band structure that exhibits systematic evolution with strain, revealing a clear correlation between local curvature and bandgap modulation.
However, generating a comprehensive DFT database for continuously variable curvature across the full range of experimentally observed deformations is computationally prohibitive. To overcome this limitation, a recurrent neural network (RNN) model is trained on a representative dataset of DFT derived strain vs density of states (DOS) relationships, enabling the prediction of an extended range of electronic properties without additional simulations. The RNN,  in contrast to standard interpolation schemes, dynamically learns the full nonlinear strain-DOS relationship, enabling prediction of spatially resolved electronic structure, i.e. the DOS including the CBM and VBM energies, and $E_g$ across arbitrary and continuously varying strain distributions. The trained RNN is superior to standard polynomial interpolation schemes, which fail to fully capture the nuances of a multiparameter training dataset.  
The model is validated by comparing predicted $E_g$ with spatially resolved photoluminescence (PL) spectroscopy data on non-uniformly strained MoS$_2$ samples, fabricated by transferring ML flakes onto periodically patterned substrates. Atomic force microscopy (AFM) measurements provide the surface morphology and corresponding strain maps, from which the trained RNN predicts the spatially resolved DOS distribution. Strong agreement between the predicted parameters, like the local $E_g$ and the experimental PL emission peak energy, confirms the reliability of the computational framework. 
In addition, the results demonstrate that biaxial bending is significantly more effective than uniaxial deformation in enhancing the dielectric response and, consequently, carrier mobility. For the same magnitude of lattice deformation, biaxial bending produces a substantial $\sim$ 7\% increase in the dielectric constant, whereas uniaxial bending yields only $\sim$ 1\%. These findings underscore biaxial bending induced strain engineering as a promising strategy for tuning dielectric screening and optimizing charge transport. 
The combined DFT-RNN-spectroscopy framework introduces a data-driven pathway for predictive engineering of strain, enabling the design of defect-free, flexible, and high-performance optoelectronic devices.
Overall, local biaxial bending provides a realistic and efficient mechanism for tuning the physical and chemical properties of TMDC monolayers, offering new opportunities for strain-adaptive optoelectronic and quantum devices.

\section{Materials and Methods}
The DFT calculations were performed using the Quantum Espresso software \cite{Giannozzi2009-jp, Giannozzi2017-vu}. The exchange–correlation interactions were treated within the generalized gradient approximation (GGA) using the Perdew-Burke-Ernzerhof (PBE) functional. Optimized norm-conserving Vanderbilt pseudopotentials, incorporating full relativistic effects and core corrections, were employed.\cite{Hamann2013} Temperature dependent DOS calculations were performed using the Python package, BoltzTraP2 \cite{madsen2018boltztrap2}. 
The monolayer MoS$_2$ was modelled using a supercell of $5\times5$ unit cells in the $xy$ plane and a 20\AA vacuum in the $z$-direction. To obtain the strained structure, the Mo atoms were confined to a 3D Gaussian surface with the centre of the supercell displaced by a specified distance $h$ from the basal plane, as shown in fig. \ref{fig:1}a. Subsequently lattice relaxation was performed for each $h$ using a $3 \times 3 \times 1$ Monkhorst-Pack K-point mesh. The kinetic energy cutoff for the plane-wave basis was 50 Ry, and the convergence threshold for atomic forces was $10^{-3}$ eV/atom. Following structural relaxation, self-consistent field (SCF) calculations were performed using a $5\times5\times1$ K-point mesh with a convergence threshold of $10^{-6}$ Ry. The DOS was then calculated using a subsequent non-self-consistent field (NSCF) calculation. Band unfolding for the supercell data was performed using the BandUP(py) code.\cite{PhysRevB.89.041407} See Supporting Information (SI) Section S1 for further details.
Energy dependence of the dielectric constant was calculated using the independent-particle approximation \cite{souza2002first}, with 156 energy bands, and a full K-space grid, under both uniaxial and biaxial bending. 
The RNN model was implemented using the Keras library with a TensorFlow backend. A one-to-many architecture was employed, consisting of seven 1D hidden layers with dimensions: 50, 500, 2000, 4000, 4000, 2000, and 700 neurons, respectively. The model was trained with 50 data sets of calculated electronic DOS at 300 K and their corresponding biaxial strain in the range 0 - 0.33\%. After 300 epochs of training, a mean squared error of \(< 0.0002\%\) was achieved. For a \% strain as input, the trained model yields the corresponding DOS. Validation of the neural network has been discussed in the SI, Section S1.5. 

The experimental investigations were conducted on chemical vapor deposition (CVD) grown MoS$_2$ ML, transferred onto SiO$_2$/Si substrates patterned with a square periodic array of cylindrical gold nanopillars of radius $\sim$ 500 nm, height $\sim$ 50 nm and periodicity $\sim$ 2.5 $\mu$m. Sample fabrication details are available in SI Section S2 \cite{kayal2023mobility}. 
The spatially resolved topography and current maps were recorded using a conductive atomic force microscopy (AFM, Bruker Multimode 8) in tapping mode using an aluminium-coated cantilever (details are available in SI section S2.3). The spatially resolved PL and Raman spectroscopy investigations were conducted at room temperature using  a confocal setup (Horiba Xplora Raman-PL system). The spatially resolved strain maps were  calculated using AFM topography by calculating the $\varepsilon_{zz}$ component of the strain tensor using continuum elasticity theory. \cite{landau1986theory}, 
\begin{equation}
    \varepsilon_{zz} = \left| \frac{\eta t}{1-\eta} \left[\frac{\partial^2 h}{\partial x^2} + \frac{\partial^2 h}{\partial y^2}\right] \right|
\label{eqn1}
\end{equation}
\noindent
where $\eta=0.25$ is the Poisson’s ratio\cite{liu2014elastic}, $t$=0.8 nm is the ML thickness, and $h$ is the local topographic height. Spatially resolved strain maps were also estimated using spatial variations in the spectral shifts of the two primary Raman modes of ML MoS$_2$. The strain maps were used in conjunction with the trained RNN model to predict spatially resolved local DOS and $E_g$ maps across a sample. Note that the spatial resolution of the strain maps generated from the diffraction-limited Raman spectroscopy maps are significantly lower than those from AFM topography. Finally, the spatially resolved variation in $E_g$ across MoS$_2$ ML is compared with the spatial variation in PL spectral peak energy map to validate the model and its predictions.

\section{Results and Discussion}
\begin{figure}
    \centering
\includegraphics[width=\linewidth]{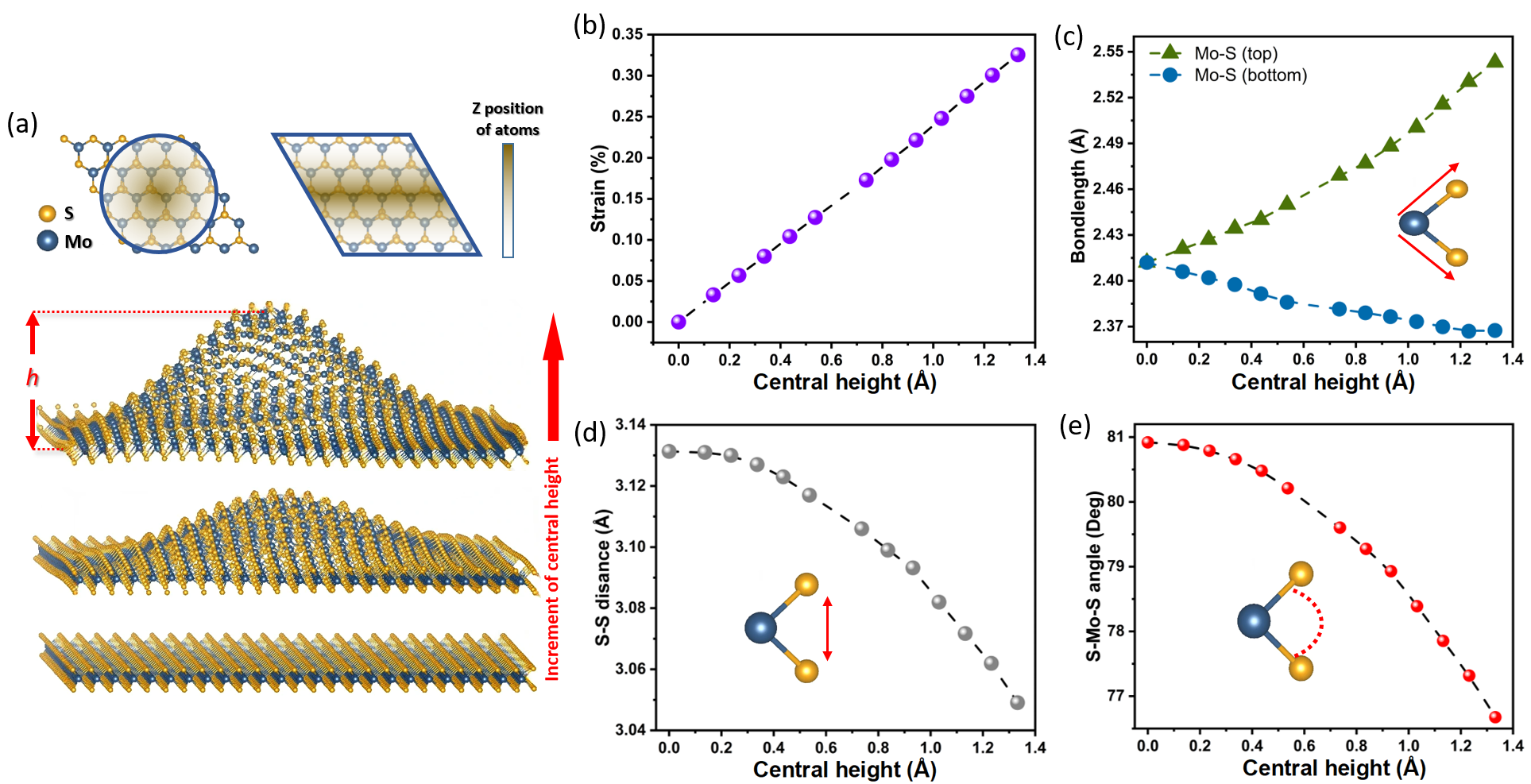}
    \caption{(a) (top) MoS$_2$ supercell used in DFT calculations, showing biaxial and uniaxial bending with the darker shades denote higher $z$ values. (bottom) Schematic of biaxially bent ML MoS$_2$ lattice with increasing $h$. Variation of (b) \%  induced strain with $h$, (c) Mo-S bond lengths, (d) S-S distance and (e) S-Mo-S bond angle with central height of the deformation.}
    \label{fig:1}
\end{figure}
Bending of the lattice induces atomic-scale structural deformations in the bond lengths and bond angles. In ML MoS$_2$, tensile strain elongates the Mo-S bond length, decreases the S-Mo-S bond angle and decreases the out-of-layer S-S separation. The deformations modulate both the phononic and electronic band structure, altering $E_g$ and at high strain transitioning from a direct to an indirect band gap system. Strain also alters the energy separation between the K and $\Gamma$ points and modifies exciton binding energies, while largely preserving the intrinsic D$_{3h}$ symmetry of the lattice. 
 Importantly, the band structure shows a significantly higher sensitivity to biaxial strain compared to uniaxial strain, as do other physical properties, including E$_g$. Evolution of the electronic band structure of ML MoS$_2$ under `in-plane' uniaxial and biaxial strain, both tensile and compressive is presented in SI Section S1.1 and summarized in Table S1. Both tensile and compressive leads to a reduction in $E_g$ with higher sensitivity to biaxial strain with the calculated $E_g\rightarrow 0$ above 10\% biaxial strain.

However, bending-induced strain, as explored here, differs significantly from in-plane strain counterparts. Fig. \ref{fig:1}a shows the deformed lattice used to calculate local strain using equation \ref{eqn1}. Here, the central height ``$h$" of the deformation is used as a single parameter characterizing the strained state of the lattice, which is a readily measured quantity, e.g. via scanning probe microscopy topography.  
For bending induced lattice deformations, the changes to the lattice parameters are spatially non-uniform, with the largest change induced at the regions of highest local strain (curvature). The resulting `local' modifications of the band structure render the stoichiometrically homogeneous ML inhomogeneous in its electronic properties. Fig. \ref{fig:1}b shows that within the explored regime, the calculated \% strain varies linearly with $h$.
Characteristically, bending simultaneously induces tensile and compressive strain in the top and bottom of a layer, which is evidenced here by the differential change in the respective Mo-S bond lengths, as shown in fig. \ref{fig:1}c. Further, the out-of-plane S-S distance decreases (fig. \ref{fig:1}d) along with a commensurate decrease in the S-Mo-S bond angle, as shown in fig. \ref{fig:1}e, effectively reducing the monolayer thickness. For all cases in figs. \ref{fig:1}(b-e) the ordinates are those calculated at the point of the highest strain (curvature) of the lattice.  
\begin{figure}[t]
    \centering
\includegraphics[width=0.8\linewidth]{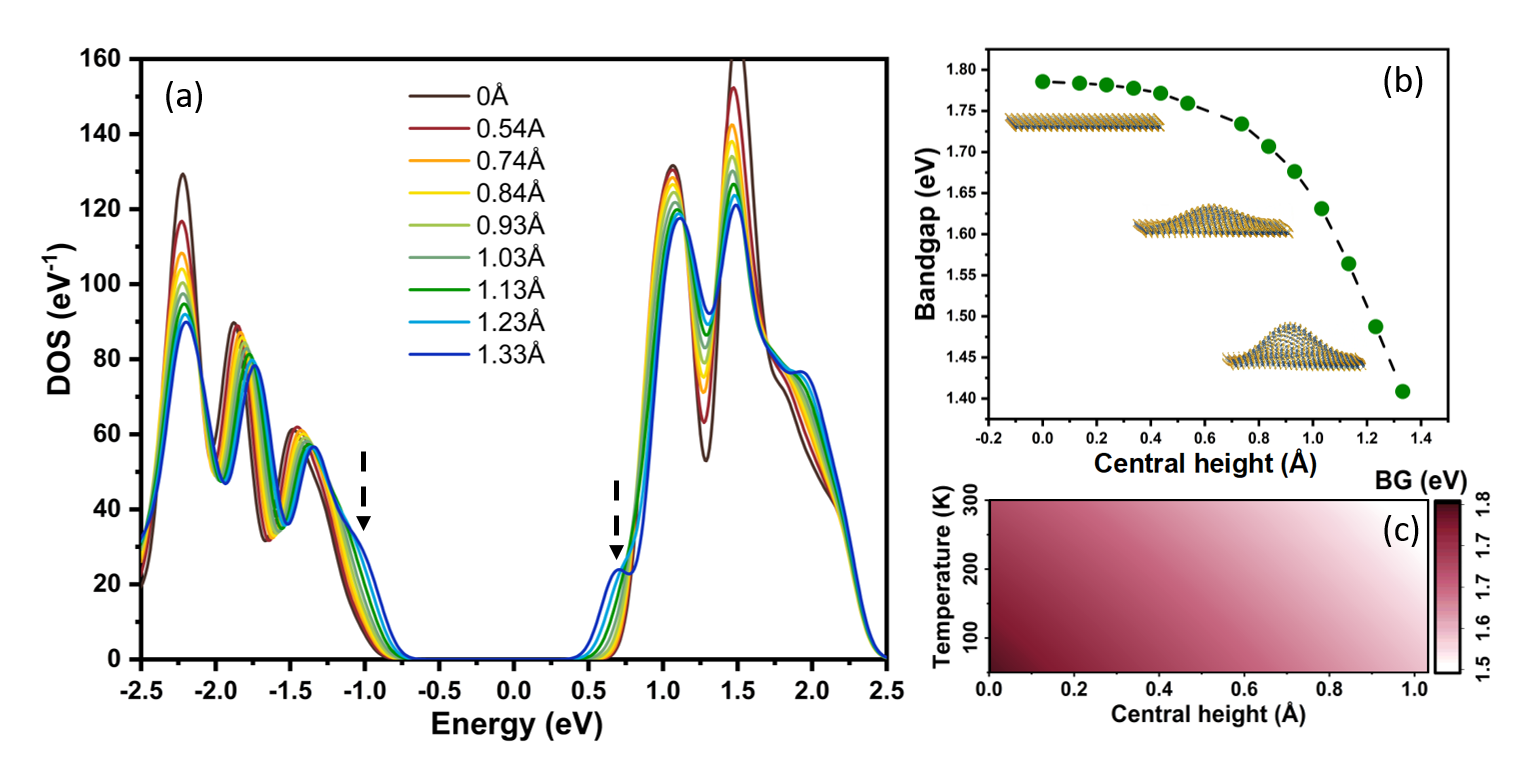}
    \caption{(a) Evolution of electronic density of states with $h$ (the arrows indicate additional band edge states due to bending), (b) variation of band gap energy ($E_g$) with $h$ and (c) variation of $E_g$ with height and temperature.}
    \label{fig:T}
\end{figure}
DFT calculations on the bent lattice show that increasing $h$, and thus the strained state of the ML, significantly affects the calculated DOS, as shown in fig. \ref{fig:T}a. Appearance of additional states at the band edges shifts the conduction band minimum (CBM) to lower energies and the valence band maximum (VBM) to higher energies, as shown in SI fig. S4(a-b). The analysis shows that, for a deformation of $h$ = 1.33 \AA, the CBM energy decreases by 25\% and increases the VBM energy by 20\%, with the E$_g$ decreasing by $\sim$22\%, as illustrated in fig. \ref{fig:T}(b). 
Notably, the band structure modifications under biaxial bending are substantially larger than those under comparable planar biaxial strain (see SI Table S1 and Fig. S1). For instance, the above biaxial bending of $h$=1.33\AA ~induces $\sim$0.35\% strain and $\sim$ 22\% reduction $E_g$, whereas a planar strain of $\sim$ 2-3\% would be required to achieve a comparable decrease in $E_g$ \cite{Yu2016-dv, Gonzalez2018-xj}. Further, finite temperature calculations predict a further decrease in the effective $E_g$ with increasing temperature, as shown in the figures. \ref{fig:T}(c). Refer SI section S1.3 for details of finite temperature calculations.
Experimentally, it has been demonstrated that lattice strain enhances carrier mobility in MoS$_2$ between 10 to 10$^3$ times \cite{ng2022improving, kayal2023mobility, Datye2022-ic}. 
Local strain features, such as wrinkles and bubbles, locally reduce the material $E_g$ and thus enhance the free-carrier density \cite{kayal2023mobility, Darlington2020-pf}, thereby improving the screening of charged impurities, reducing scattering, and increasing carrier mobility. 

\begin{figure}[t]
    \centering
\includegraphics[width=1\linewidth]{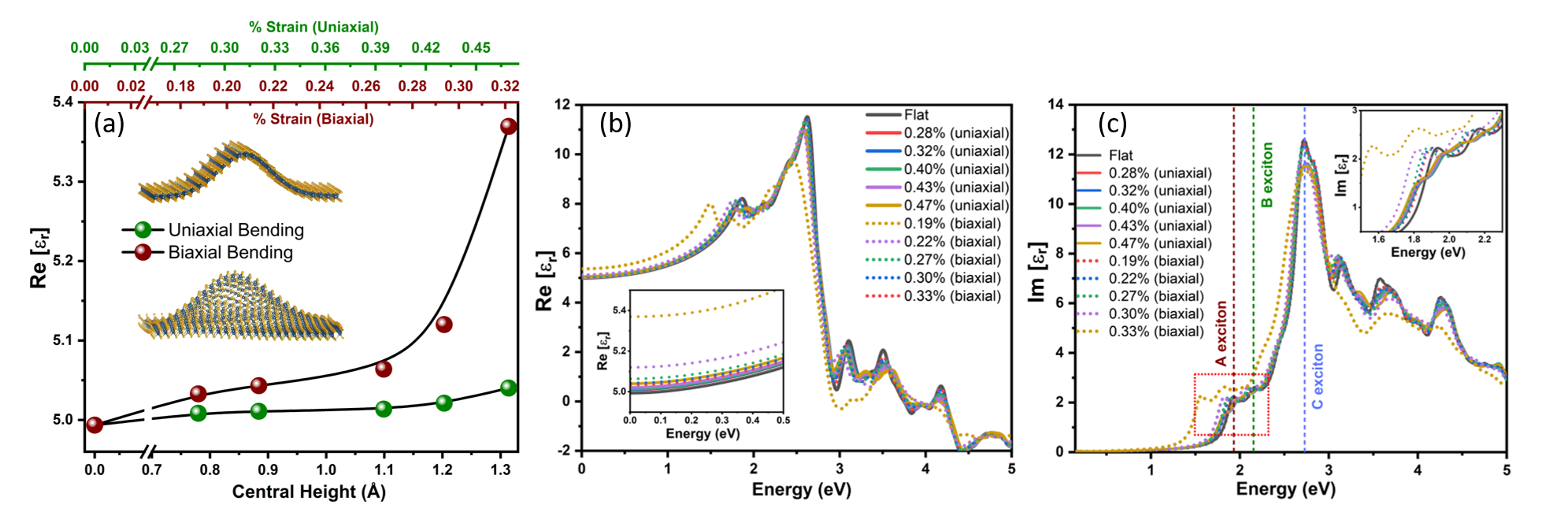}
    \caption{(a) Variation of calculated dc in-plane $\varepsilon_r$ of ML MoS$_2$ with deformation $h$ under biaxial and uniaxial bending. (b) Energy dependence of in-plane $\varepsilon_r$ for flat and biaxially and uniaxially bent ML. Inset shows a zoomed view of $\varepsilon_r$ close to zero energy. (c) Energy dependence of in-plane $\varepsilon_i$ for flat and biaxially and uniaxially bent ML, zoomed view of the marked area has shown in the inset}
    \label{fig:4}
\end{figure}

Bending of the MoS$_2$ ML breaks in-plane lattice symmetry and softens the phonon modes (E$_{2g}$ and A$_{1g}$), enhancing Born effective charges and softening polar vibrations, which collectively increase the dielectric constant ($\varepsilon_r$) through stronger ionic polarization. Fig. \ref{fig:4}(a) plots the change in the in-plane dc Re[$\varepsilon_r$] under biaxial and uniaxial bending of the ML. The top axes show the corresponding maximum induced strain, and the insets depict the lattice deformation in either case. The in-plane dc Re[$\varepsilon_r$] increases by $\sim$ 7\% under biaxial and by $\sim$ 1\% under uniaxial bending under comparable deformation $h$=1.3\AA, as shown in fig. \ref{fig:4}(a). Fig. \ref{fig:4}(b-c) plot the variation of in-plane Re[$\varepsilon_r(\omega)$] and Im[$\varepsilon_r(\omega)$] with energy ($\hbar\omega$) under biaxial and uniaxial bending, for various \% strain. The plot evidences the A, B and C excitonic absorptions (dashed lines in fig. \ref{fig:4}c), which redshift under increasing strain, commensurate with the decrease in $E_g$. The results reaffirm the higher sensitivity of electronic properties to biaxial bending compared to uniaxial bending.
In summary, both the increased carrier density and the enhanced Re[$\varepsilon_r$] increase screening of long-range interactions (ionised impurities etc.) that limit electrical transport across MoS$_2$, thereby reducing scattering and enhancing mobility.  

\begin{figure}[t]
    \centering
    \includegraphics[width=0.8\textwidth]{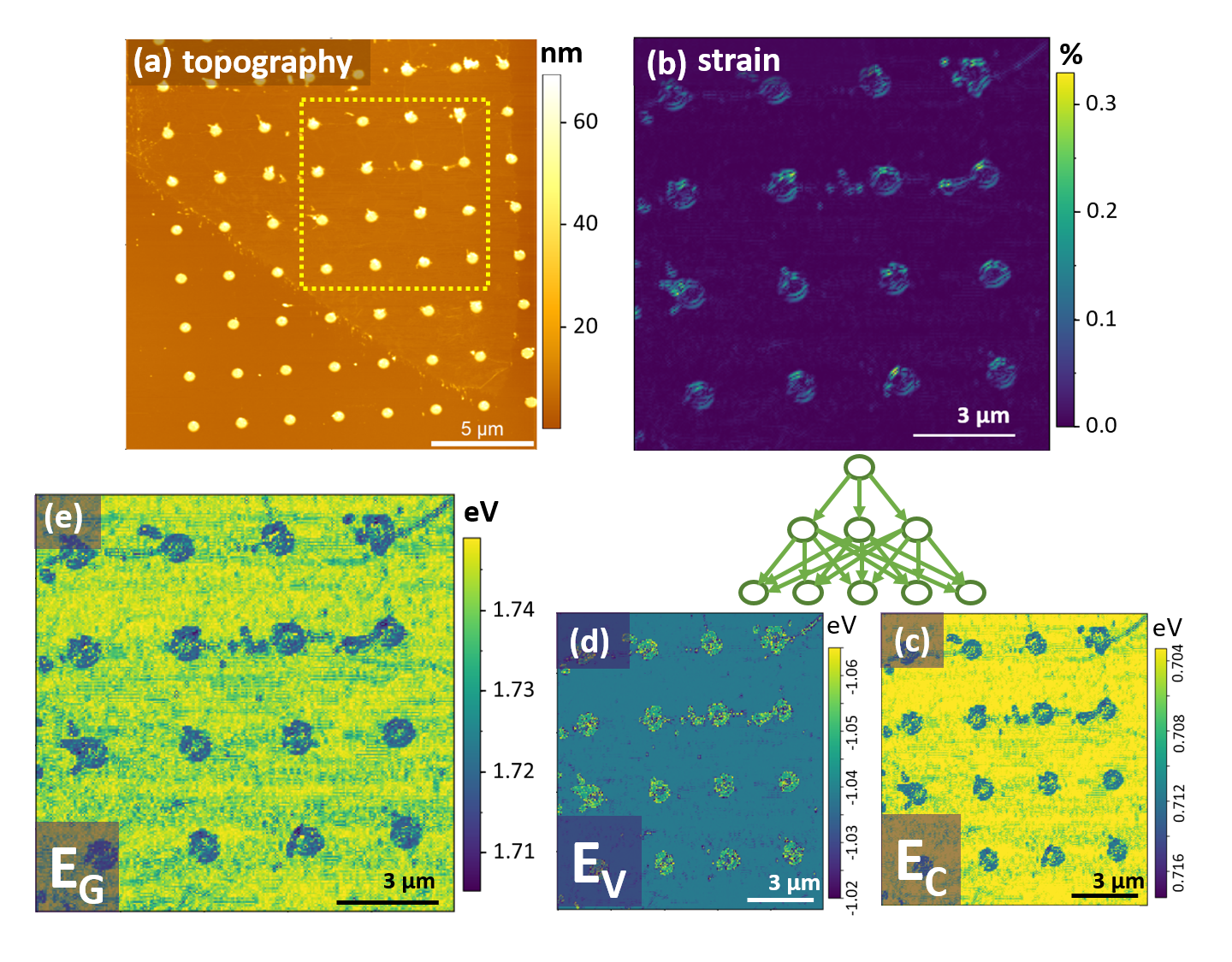}
    \caption{Steps for generating the DOS map, and followed by generating band gap map of ML MoS$_2$ on a patterned substrate (a) AFM topography of ML MoS$_2$ flake draped over gold pillars on Si, (b) Strain map, calculated from AFM topography for the selected region. Prediction map: (c) conduction band minima (CBM), (d) valence band maxima (VBM), (e) bandgap energy}
    \label{fig:pred_map}
\end{figure}

The calculated bending induced \%strain and the corresponding DOS (fig. \ref{fig:T}a) obtained from lattice deformations were used to train an RNN to predict the DOS and $E_g$ for arbitrary values of \%strain. Experimental data on non-uniformly strained MoS$_2$ is obtained from ML flakes transferred onto substrates patterned with periodic cylindrical nanostructures \cite{kayal2023mobility}.  
Topographic maps (fig. \ref{fig:pred_map}a) show that the ML conformally drapes around the cylindrical nanostructures, nucleating wrinkles and bubbles across the entire ML.
Since strain is determined by local surface curvature (equation \ref{eqn1}), high strain is generated at the edges of the pillars and at the corrugations across the flake. 
A typical example of a topography derived strain map is shown in fig. \ref{fig:pred_map}b that records spatially non-uniform strain varying between 0 - 0.35\%. The RNN uses the strain maps to predict spatially resolved local DOS along with the CBM and VBM energy maps (figs. \ref{fig:pred_map}c-d) and the $E_g$ map, as shown in fig. \ref{fig:pred_map}e. The reduction in $E_g$, primarily at the edges of the nanostructures and at the deformations (wrinkles and nanobubbles), compared to that at the flat regions, evidences the energy landscape across the ML. Variation of $E_g$ across wrinkles and bubbles, using AFM topography generated strain maps, are discussed later. 

\begin{figure}[t]
    \centering    \includegraphics[width=\linewidth]{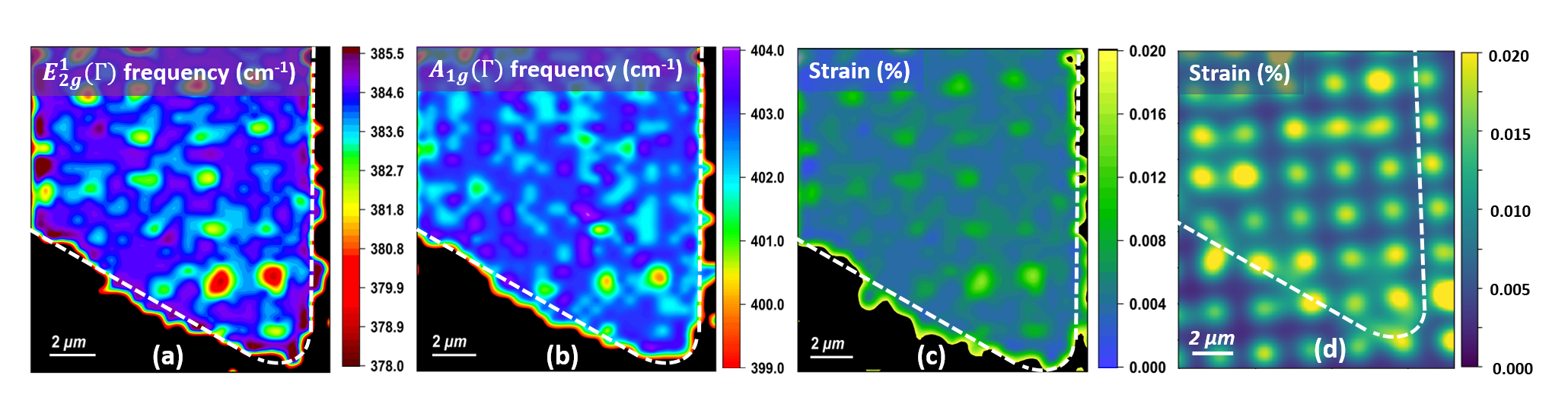}
    \caption{ Spatial maps of frequency for (a) \(E_{2g}^{1}\) ($\Gamma$) and (b)\(A_{1g}\) ($\Gamma$) Raman modes of the ML MoS$_2$ flake on patterned substrate (Fig. \ref{fig:pred_map}) (c) strain map calculated from Raman spectral shifts and (d) strain map from AFM topography (white dashed line denote flake boundary)}
    \label{fig:raman}
\end{figure}

\begin{figure}[!ht]
    \centering
\includegraphics[width=\linewidth]{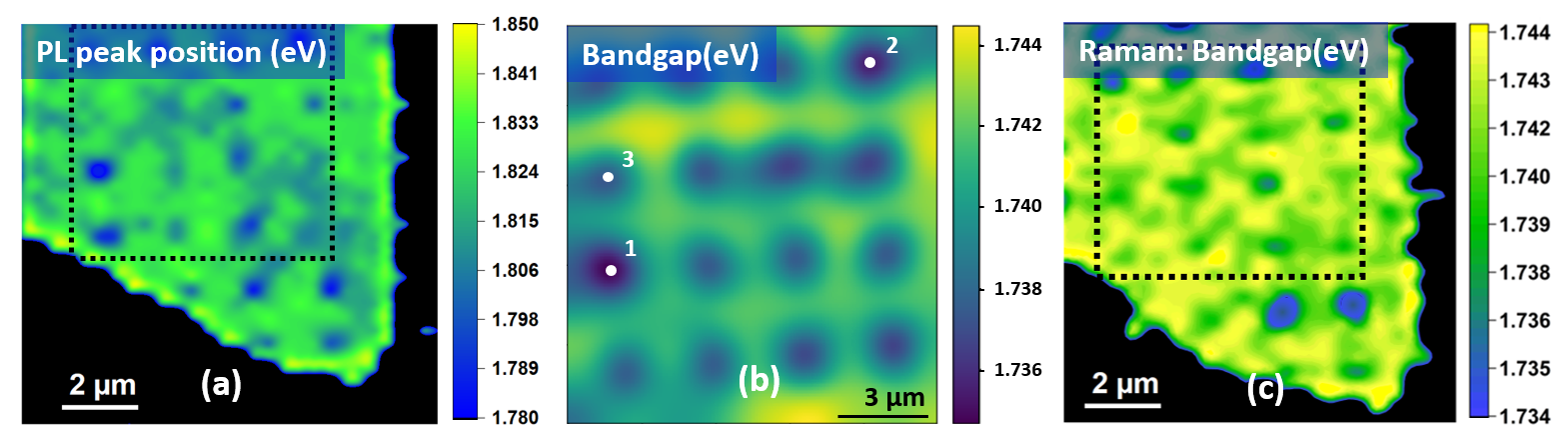}
    \caption{(a) PL map peak emission energy of ML MoS$_2$ flake on a patterned substrate, (b) spatial variation of E$_g$ calculated from topography generated strain map (c) spatial variation of E$_g$ calculated from strain calculated from Raman spectral shifts. The black dotted line demarcates the region used for the AFM topography.}
    \label{fig:PL}
\end{figure}

To test the efficacy of the above methodologies, both for estimating spatial variation in strain and for predicting local $E_g$ using the trained RNN, complementary techniques such as spatially resolved Raman and photoluminescence spectroscopy maps were employed. 
Fig. \ref{fig:raman}(a) and (b) maps the Raman peak shifts corresponding to the \(E_{2g}^{1}\) and \(A_{1g}\) modes across an ML MoS$_2$ on a patterned substrate. The change in peak shifts of the two modes, with respect to those from a flat, unstrained MoS$_2$ (SI fig. S11) was used to calculate a strain map as shown in fig. \ref{fig:raman}c \cite{zhang2024enhancing}. (See SI section S2.4 for further details.) The strain map generated from the Raman spectral map shows high correlation with that generated from the AFM topography, as shown in Fig. \ref{fig:raman}d.   
 A direct estimate of the spatial variation of $E_{g}$ across the strained ML is obtained from the spatially resolved PL peak energy map for the A-exciton, as shown in fig. \ref{fig:PL}a. 
 Corresponding to the region enclosed by the dashed box in fig. \ref{fig:PL}a, the spatial variation of the local $E_g$, predicted by the RNN model from the strain map derived from the topography, is shown in fig. \ref{fig:PL}b. Similarly, the $E_g$ map predicted from the Raman spectral shift maps derived strain map is shown in fig. \ref{fig:PL}c.  
Atop the nanostructures, the PL map shows that the emission peak energy is `red-shifted' compared to that from an unstrained flat region, consistent with the reduced $E_g$ predicted by the RNN trained on the DFT results. Although the maps show good qualitative correlation between experiment and simulations, there are noticeable exceptions, as in the case of the region denoted by point 3 in fig. \ref{fig:PL}b. The predicted low $E_g$ region around the point does not correspond to a lower PL peak energy or lower predicted $E_g$ from the Raman derived strain map. Upon closer inspection, the topography shows that the flake in the region is perforated by the nanostructure, and the lattice is thus locally unstrained, as discussed in SI section S1.5. 
It is worth noting that the above DFT calculations employing semilocal functionals (GGA) systematically underestimate $E_g$ by $\sim$ 50 meV compared to those from optical spectroscopy. It arises from the absence of derivative discontinuity in the exchange-correlation energy, causing Kohn-Sham eigenvalues to poorly approximate quasiparticle energies.\cite{Morales-Garcia2017-gn, Bagayoko2014-uz, Bystrom2024-kv} Further, a AFM-derived strain map affords substantially higher spatial resolution than strain maps obtained from optical spectroscopy. 
Thus, the former enables a more detailed correlation between local deformations and electronic properties than the latter, which is compromised by spatial averaging, leading to smaller estimates of strain and thus $E_g$ variation.    
Overall, the spatial variation of $E_g$ predicted by the model at the wrinkles and nanobubbles shows strong agreement with the spatial variation in PL peak energy, confirming that the framework faithfully captures strain induced electronic structure modulation. 

\begin{figure}[!ht]
    \centering    \includegraphics[width=\linewidth]{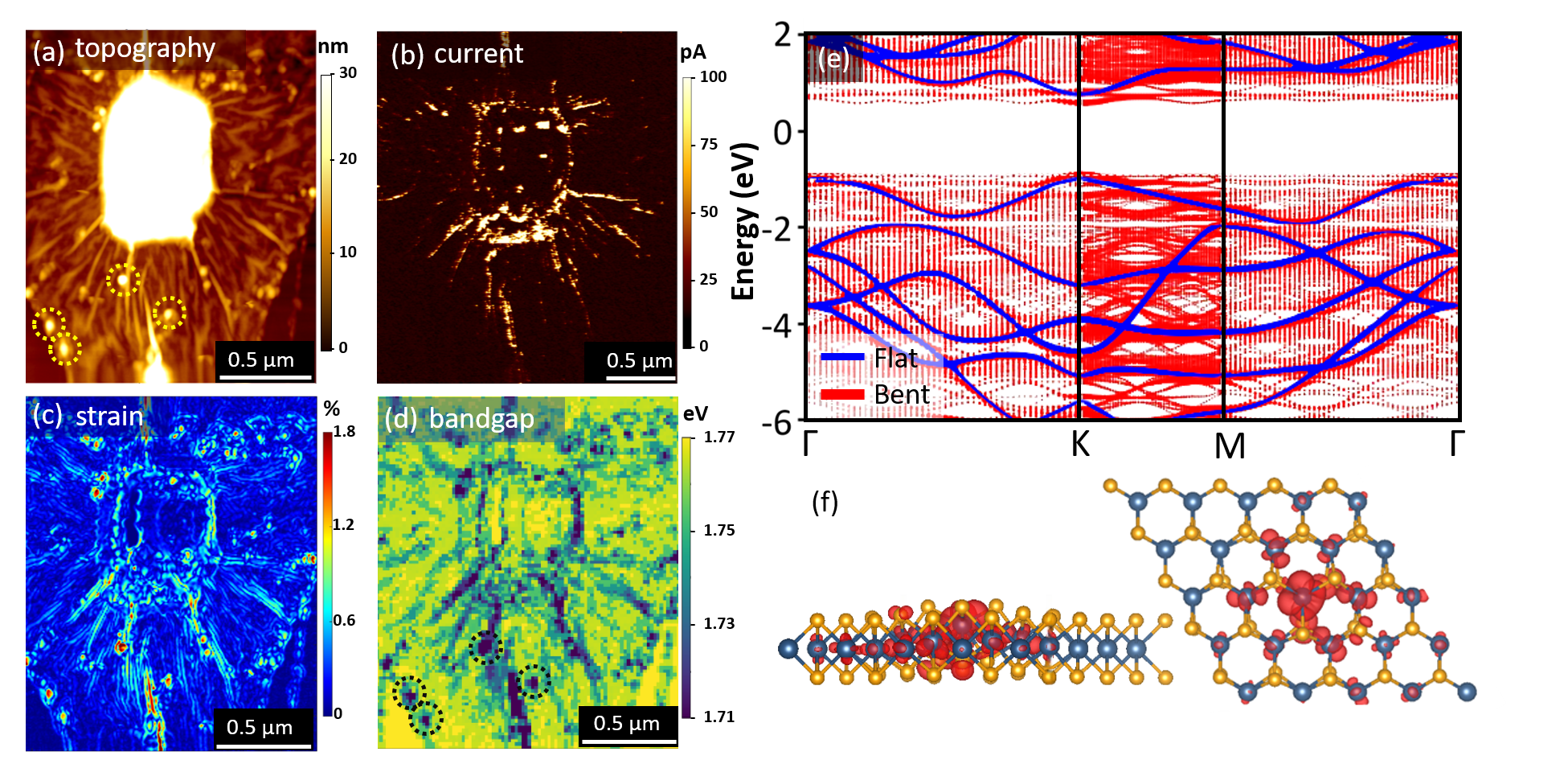}
    \caption{(a) AFM topography of MoS$_2$ ML draped over a nanopillar and the corresponding (b)current map at +0.2V sample bias (c)  strain map (d) RNN predicted bandgap map. Dotted circles indicate the location of nanobubbles (e) unfolded electronic band structure of flat (blue) overlaid on that of biaxially bent (red) MoS$_2$ ML with $h$=1.33 \AA. (f) side and top view of the charge density iso-surface corresponding to the CB edge state at $\sim 0.7 eV$, showing charge density - strain correlation. }
    \label{fig:ChargeLoc}
\end{figure}

For an MoS$_2$ ML draped over a nanopillar, fig. \ref{fig:ChargeLoc}(a) shows the AFM topography revealing a network of wrinkles around the nanostructure along with nanobubbles (marked by circles), arising from non-uniform strain imposed by the underlying nanostructure. The simultaneously acquired current map provides direct experimental evidence for strain induced charge localization as shown in fig. \ref{fig:ChargeLoc}(b). The current map resolves the high-strain features as regions of significantly higher current, indicative of higher local conductance and hence increased local charge density.
The higher spatial resolution is translated to the topography derived strain map illustrated in fig. \ref{fig:ChargeLoc}(c), which was used to calculate the spatial distribution of $E_g$ shown in fig. \ref{fig:ChargeLoc}(d). Note that to reduce the computational cost of calculating $E_g$ map via the RNN, the spatial resolution of the strain map in fig. \ref{fig:ChargeLoc} (c) was reduced from $512\times512$ to $128\times128$ via local averaging, which also reduces the upper limit of the strain range to below 0.5\%.  
Significantly, this spatial concurrence of wrinkles and nanobubbles (high strain) with high current and lower $E_g$ is fully consistent with the DFT predicted band structure and charge density localization at regions of maximum curvature (strain). 
Fig. \ref{fig:ChargeLoc}(e) shows the calculated band structure for the biaxially bent supercell (fig. \ref{fig:1}(a)) in red, overlaid with that of the flat unstrained lattice in blue, which reveals additional spectral features, especially evident at the band edges (K - M points).
The calculated DOS evidences these band edge states in fig. \ref{fig:T}(a) and fig. S7, which yields a reduced $E_g$ at the strained regions. 
These strain induced band-edge states introduce spatial charge density inhomogeneities in the lattice despite the stoichiometric uniformity of the ML. Fig. \ref{fig:ChargeLoc}(f) presents the partial charge isosurface corresponding to the CBM edge states of the biaxially bent lattice, revealing charge localization at the centre of the supercell, where curvature and strain are highest, in direct agreement with the experimentally observed elevated current at wrinkles and nanobubbles. The evolution of this charge localization with increasing deformation height is discussed in SI Section S1.7. The results are also consistent with the variation in Raman spectral shifts between strained and unstrained regions, which reflects locally varying electron density \cite{Yang2014-qg, zhang2024enhancing}.  
The sub-diffraction spatial resolution of the AFM data derived maps thus provides a critical experimental validation of the strain–charge density correlation predicted by the computational framework, at length scales inaccessible to optical probes. 

Collectively, these results demonstrate the overall capability of the framework to quantitatively model the electronic structure of nonuniformly strained 2D systems using a practical, computationally efficient approach across varied strain configurations. 
Further, for impurity scattering dominated transport in CVD grown TMDCs, these biaxially bent deformations enhance $\varepsilon_r$ along with the charge density, which together suppress Coulomb scattering, resulting in significant improvements in carrier mobility, as reported experimentally \cite{ng2022improving,kayal2023mobility}. Given that wrinkles and nanobubbles are an intrinsic and often unavoidable feature of such device geometries, accurately capturing their electronic and phononic consequences and even exploiting their incidence is essential for the rational design and engineering of high-performance TMDC based devices.

\section{Conclusion}
In conclusion, we have investigated local changes to the electronic properties of ML MoS$_2$ to non-uniform biaxial bending by combining DFT calculations, an RNN framework, and spatially resolved experimental characterization. The results show that biaxial bending induced strain is substantially more effective than either uniaxial bending or in-plane strain in modifying electronic and dielectric properties. A $\sim$ 0.35\% strain induced by biaxial bending results in $\sim$ 22\% reduction in band gap and $\sim$ 7\% increase in dielectric constant, compared to a $\sim$ 5\% reduction in $E_g$ and $\sim$ 1\% increase in dielectric constant under comparable uniaxial bending. 
The reduction in band gap may be traced to the strain induced change in electronic band structure that introduces band edge states, concentrating carriers in regions of high strain (curvature), effectively increasing the local free electron density. This is consistent with the enhanced electrical conductance observed experimentally at the high strain features like wrinkles and nanobubbles.
The RNN trained on DFT calculated DOS of biaxially bent ML MoS$_2$ lattice enables rapid, spatially resolved prediction of DOS and band gap maps directly from AFM topography, eliminating the computationally intensive cost of DFT calculations at all points across samples with continuous strain distributions. 
The predicted band gap maps show good agreement with spatially resolved PL peak energies, with systematic deviations attributable to known theoretical and experimental limitations. 
Crucially, wrinkles and nanobubbles, unavoidable features of 2D systems, are shown to locally decrease the band gap, and increases local charge density and dielectric screening, which positively impacts electrical transport in MoS$_2$. The DFT-RNN-spectroscopy framework developed here provides a computationally efficient and validated route towards predictive strain engineering in 2D systems and is directly extensible to other TMDCs and van der Waals heterostructures. The results offer a foundation for the design of strain-adaptive, flexible optoelectronic and nanoelectronic devices.
These results highlight the importance of strain symmetry in tuning the electronic behavior of transition metal dichalcogenides and establish a practical framework for analyzing strain-induced property modulation in realistic device geometries. The approach developed here may be extended to other two-dimensional semiconductors and offers a pathway toward strain-engineered flexible optoelectronic and nanoelectronic devices.

\section{Data and Code Availability}
Data supporting the findings of this study and the RNN codes used are available from the corresponding authors on reasonable request.

\begin{acknowledgement}
Authors acknowledge Prof. M M Shaijumon
(IISER Thiruvananthapuram) for the use of CVD growth facilities. Authors acknowledge financial support from ANRF, Government of India (CRG/2023/006878), MOE-STARS (STARS-2/2023-1012), SPARC (No. 3086), and IISER Thiruvananthapuram for computing time on the Padmanabha cluster. SD acknowledges a PhD fellowship from DST INSPIRE. RN acknowledges University Grants Commission, Govt. of India, for a
PhD fellowship. 
\end{acknowledgement}
\begin{suppinfo}
The supporting information file is available at.
It includes details on the construction of the bent 2D lattice, the calculated electronic band structures of variously strained lattices, finite-temperature extrapolation of the calculated parameters, neural network modelling details, and a discussion of dielectric constant estimation for 2D systems. Experimental details of sample fabrication and characterisation are included and referred to in the manuscript.
\end{suppinfo}
\clearpage
\appendix
\renewcommand{\thesection}{S\arabic{section}}
\setcounter{section}{0}

\section*{Supporting Information (SI)}
\addcontentsline{toc}{section}{Supporting Information (SI)}

\renewcommand{\thesection}{S\arabic{section}}
\renewcommand{\thefigure}{S\arabic{figure}}
\renewcommand{\thetable}{S\arabic{table}}


\begin{spacing}{1}

\section{Calculating Band Structure under Strain}
\subsection{Planar Uniaxial and Biaxial Strain}
\begin{figure}[t]
    \centering
    \includegraphics[width=1\linewidth]{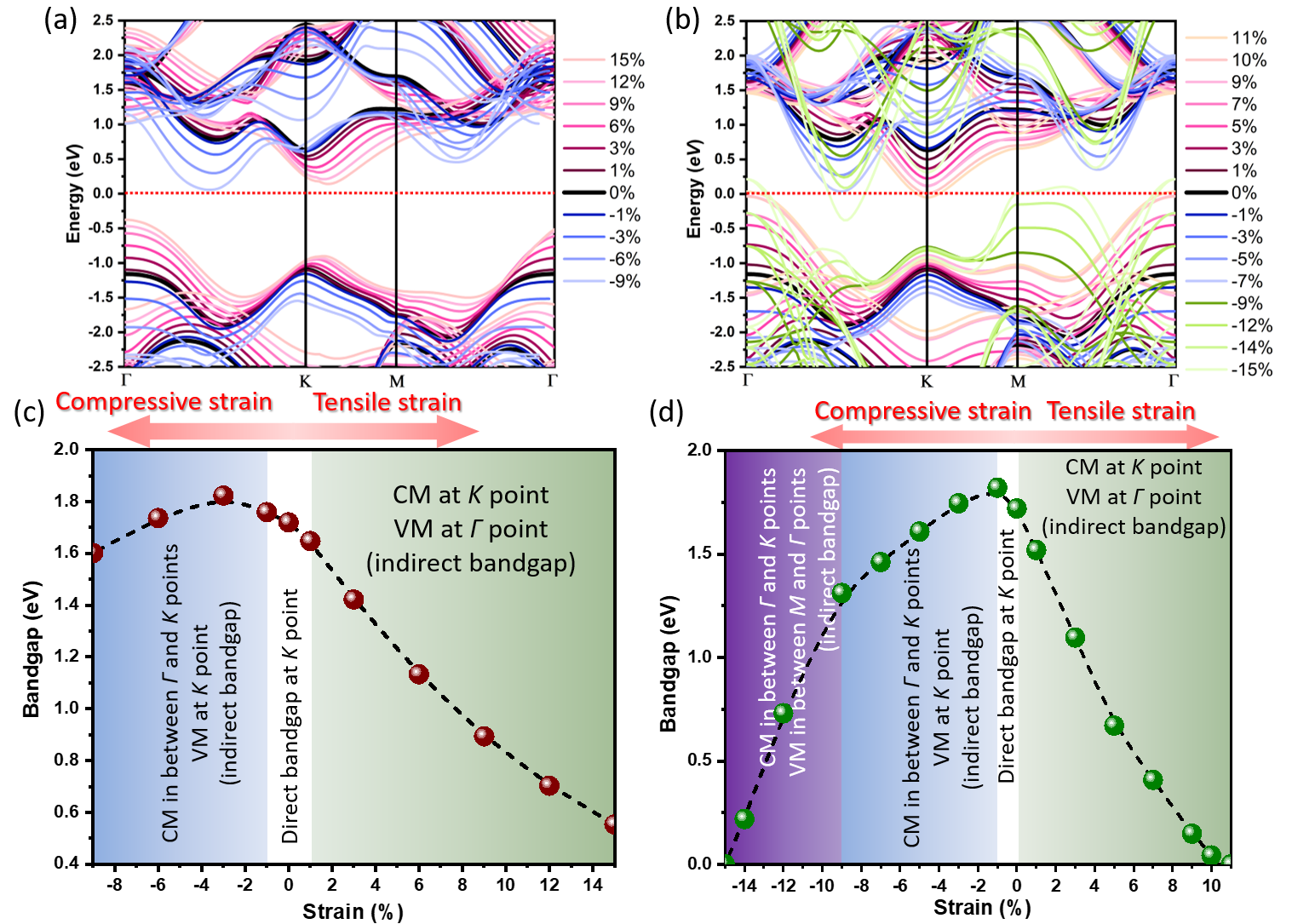}
    \caption{Evolution of the electronic band structure of MoS$_2$ under (a) uniaxial and (b) biaxial strain. Variation of the band gap energy with strain under (c) uniaxial and (d) biaxial strain. Colored bands denote positions of the CBM and VBM in the $k$-space. }
    \label{fig:6}
\end{figure}

A unit cell of MoS$_2$ was subjected to planar strain by systematically changing the Mo-Mo distance\cite{wang2022first, Mohammad_Tabatabaei2013-jb}. Change in the average Mo-S bond length along a single direction is considered to be a uniaxial strain, and along two perpendicular directions is denoted as biaxial strain, which can be both compressive and tensile. 
Figure \ref{fig:6}(a-b) shows the evolution of the calculated band structure of a monolayer MoS$_2$ for both uniaxial and biaxial strain, from compressive to tensile. The calculated band structure of the unstrained system is commensurate with previous reports\cite{Mohammad_Tabatabaei2013-jb, Zollner2019-sk, PhysRevB.83.245213}. Increasing strain, both compressive and tensile, decreases the band gap energy and renders it indirect \cite{feng2012strain, Mohammad_Tabatabaei2013-jb, chong2021first, yun2012thickness}. However, biaxial strain (fig. \ref{fig:6}d) shows a larger fractional change in band gap compared to uniaxial strain (fig. \ref{fig:6}c). Indeed, calculations show that large biaxial strain tends to reduce the band gap energy towards zero, provided the material can sustain the breaking strain of $\sim 12\%$.

\begin{table}[h!]
\centering
\begin{tabular}{ |p{2cm}|p{1.5cm}|p{2.5cm}|p{2.5cm}|p{3.2cm}| }
\hline
Strain&\% strain & CBM point & VBM point & variation in E$_g$(eV) \\
\hline
\multirow{1}{4em}{uniaxial tensile} & 0 - 1 & K & K & $1.72-1.64$\\
& 1 - 15 & K & $\Gamma$ & $1.64-0$\\
\hline
\multirow{1}{4em}{uniaxial compressive} & 0 - 1 & K & K & $1.72-1.82$\\
& 1 - 9 & $\Gamma$ & K & $1.82-1.60$\\
\hline
\multirow{1}{4em}{biaxial tensile} & 0 - 11 & K & $\Gamma$ & $1.72-0$\\
& & & &\\
\hline
\multirow{1}{4em}{biaxial compressive} & 0 - 1 & K & K & $1.72-1.82$\\
& 1 - 9 & $\Gamma$ & K & $1.82-1.31$\\
& 9 - 15 & $\Gamma$-K & M-$\Gamma$ & $1.31-0$\\
\hline
\end{tabular}
\caption{Variation in electronic band structure under in-plane strain. CBM: Conduction band minimum, VBM: Valence band maximum.}
\label{table:1}
\end{table}

\subsection{Bending Induced Strain}
\subsubsection{Biaxial bending}
\begin{figure}[!ht]
    \centering
    \includegraphics[width=0.9\linewidth]{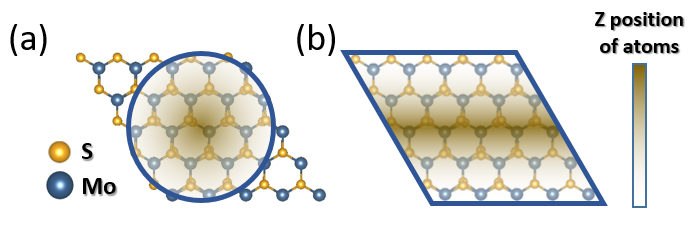}
    \caption{Schematic of the ``gaussian'' shaped bending of monolayer \(MoS_2\), here the darker regions denote elevation from the plane i.e. higher $z$ value of $Mo$, and lighter shades denote lower value of $z$. (a) biaxial bending (b) uniaxial bending}
    \label{fig:7}
\end{figure}
A $5\times5\times1$ supercell of \(MoS_2\) was used to simulate the bending induced strained lattice. 
Mo atoms outside the region denoted by the blue circle in fig. \ref{fig:7}a had their $z$ coordinate fixed on a plane (material adhesion to the substrate). The coordinates of Mo atoms inside the circle were set on the surface of a 3-dimensional Gaussian surface of standard deviation 
3.87~\text{\AA} Subsequently, the lattice was allowed to relax, with the top-most (central) Mo atom restricted to move only in the $x-y$ plane. Post relaxation, the structure departs from the initial Gaussian shape and the $z$ coordinate of the central Mo atom defines the height ($h$) of the deformation. Band structure calculations were performed for various values of $h$ of the deformed lattice, which correlates with the bending-induced \% of strain in the monolayer. 

\subsubsection{Uniaxial bending}
For simulating uniaxial bending the Mo atoms situated on the central zig-zag direction (blue line in fig. \ref{fig:7}b) were elevated to a height $h$ and the edge Mo atoms lay on a plane. Constraining the $z$ coordinate of these two sets of Mo atoms the lattice was allowed to relax. Post relaxation the lowest energy configuration yielded the uni-axially bent structure. Again band structure calculations were performed for various values of $h$ of
the deformed lattice, which correlates with the bending-induced \% of strain in the monolayer.

\subsection{Estimating Strain for the bent structures}
In the case of bending induced strain in the lattice, neither the Mo-S bond length nor the bond angle are uniformly perturbed across the entire super cell, but varies with position\cite{yu2016bending}. Here, post-relaxtion, with the energy minimised lattice, continuum elasticity theory (Equation 1 of main text) was used to calculate the position dependent \% strain across the material. The average value of \% strain has been considered as the effective strain in the system.
\subsection{Temperature-dependent DOS calculation}
\begin{figure}[!ht]
    \centering
    \includegraphics[width=0.85\linewidth]{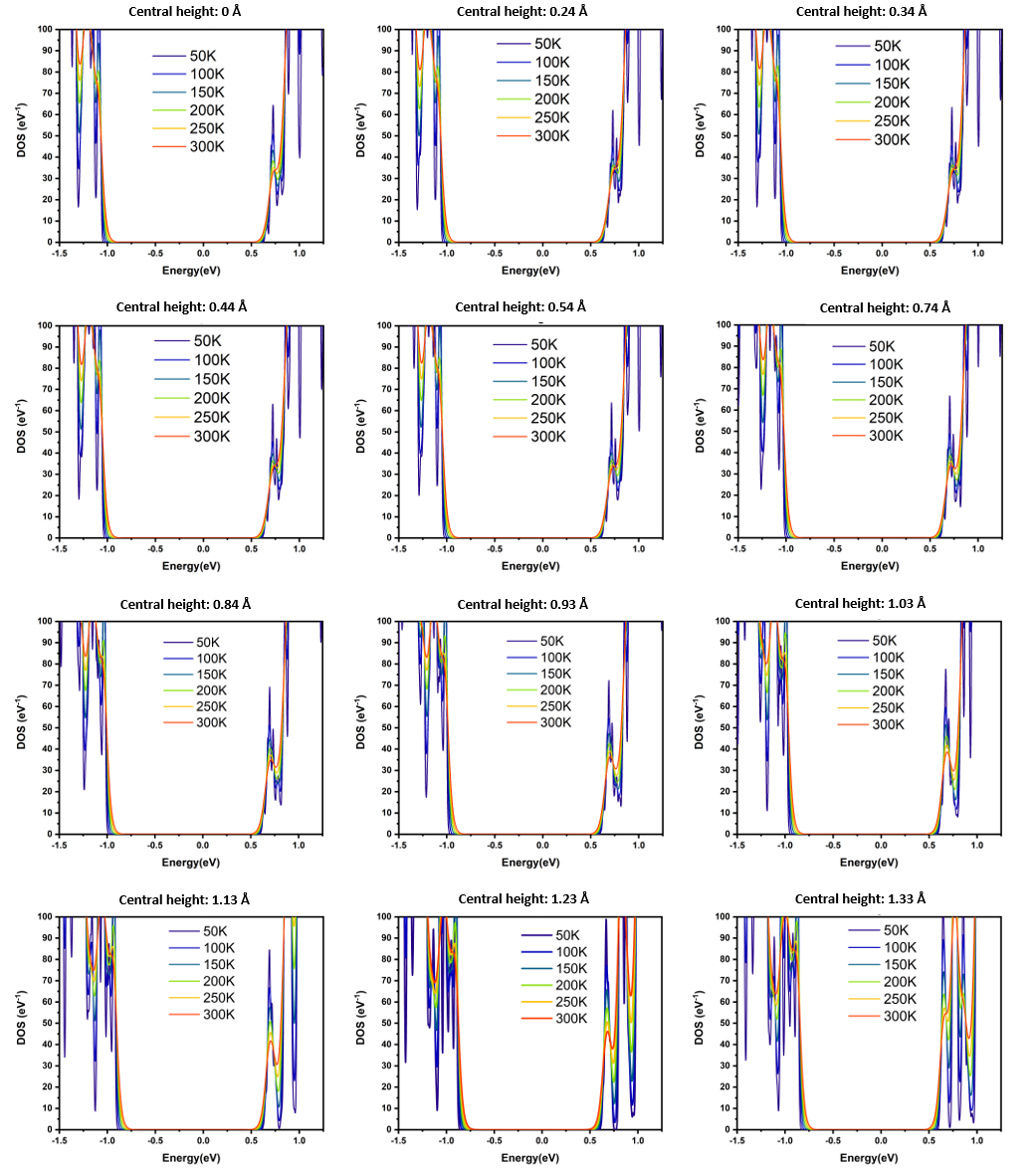}
    \caption{Temperature-dependent DOS for various values of the height $h$ of the biaxially bent lattice.}
    \label{fig:8}
\end{figure}
\begin{figure}
    \centering    \includegraphics[width=\linewidth]{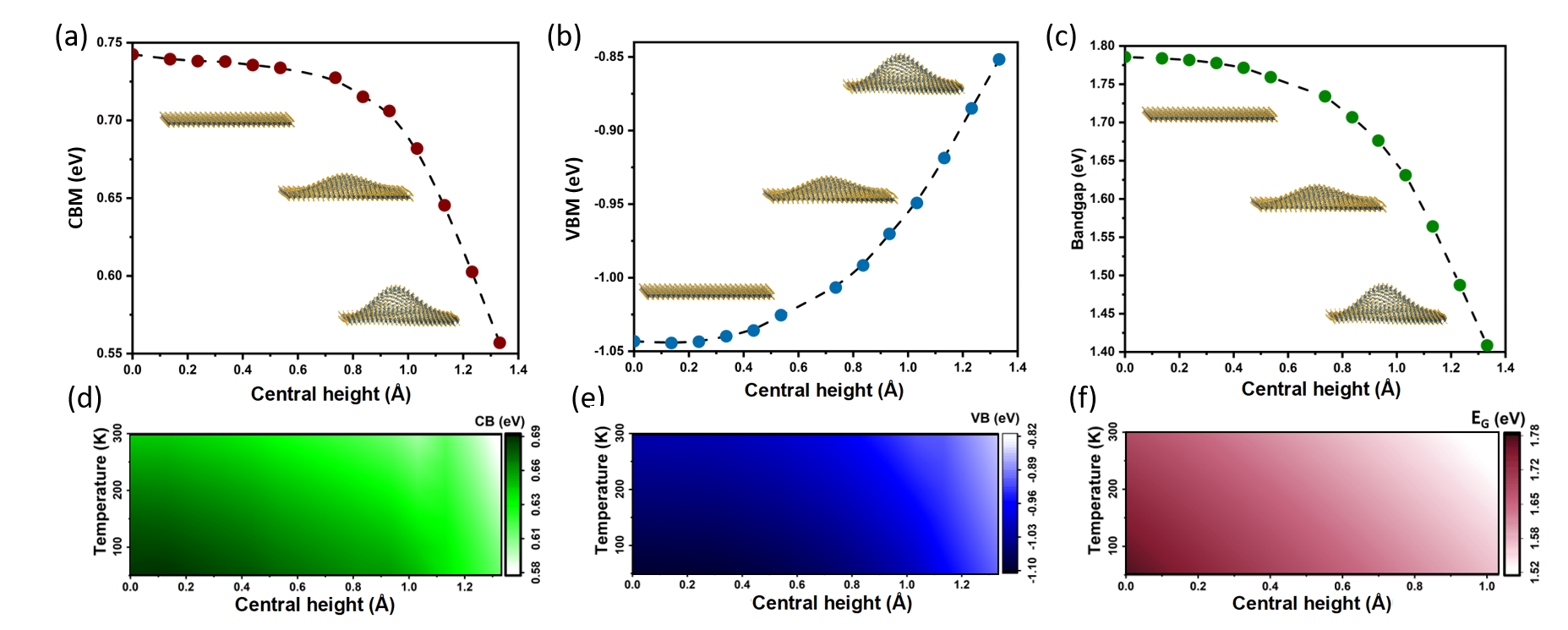}
    \caption{(a \& b) Variation of the conduction band minima (CBM) and the Valence band maxima (VBM) with respect to the Fermi level as a function of the height of deformation $h$, (c) variation of the bandgap (E$_G$) energy with $h$. (d-f) The temperature-dependent evolution of the calculated CBM, VBM, and E$_G$ with $h$.}
    \label{fig:8a}
\end{figure}
Temperature modifies the kinetic energy of electrons and lattice atoms, leading to redistribution of electronic states and broadening of energy bands\cite{gulyamov2015temperature, delice2024temperature}. The temperature dependence of the electronic band structure  were evaluated using the BoltzTraP2 package\cite{madsen2018boltztrap2}, which interpolates the electronic band structure obtained from a dense and uniform k-mesh and performs the required Brillouin-zone integrations. The resulting temperature-dependent density of states (DOS) for different values of deformation height $h$ are shown in Fig. \ref{fig:8}.  
The variation of the estimated band edge and band gap energies of the strained MoS$_2$ lattice under biaxial bending quantified via the height $h$ is shown in figs. \ref{fig:8a}(a-c). Furthermore, figs. \ref{fig:8a}(d-f) map their evolution as a function of temperature and $h$.

\subsection{Neural network modeling and prediction}
\begin{figure}[!ht]
    \centering
    \includegraphics[width=\linewidth]{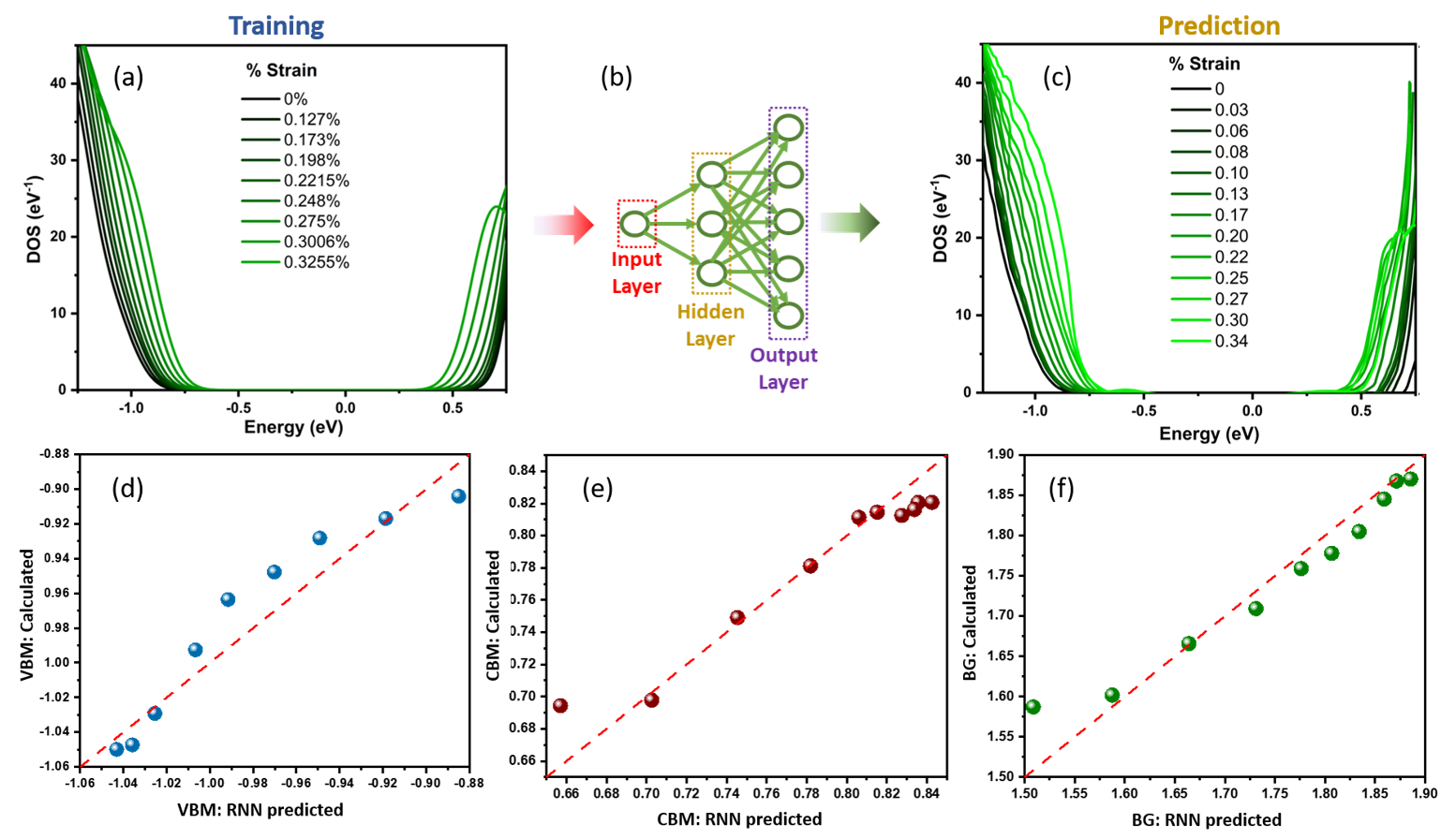}
    \caption{(a) The portion of DOS curves which have been used for the training, (b) a schematic of the neural network, (c) co-plot of predicted DOS along with two calculated DOS curves (d) the VB edge, (e) CB edge position, and (f) the BG value form the predicted DOS curve}
    \label{fig:9}
\end{figure}

The neural network approach significantly reduces computational cost. In this study, a Recurrent Neural Network (RNN) in "one-to-many" mode is used to link \% strain values to their corresponding DOS curves. Given the limited input (a single \% strain value), a deep architecture with hidden layers of $50, 500, 2000, 4000, 4000, 2000, 700$ neurons was used, respectively. Training was focused on a limited DOS region from -1.25 to 0.75 eV (fig. \ref{fig:9}a), yielding an output layer dimension of 208. After 300 epochs, the model achieved a mean squared error (MSE) of \(<0.0002\%\), enabling reliable prediction (notably, 50 datasets were used in the training, so furter improvements are possible with large training datasets). Fig. \ref{fig:9}b is a schematic of a typical neural network model. Figure \ref{fig:9}c shows predicted DOS curves for several untrained \% strain values. Notably, strain induced features such as the local maximum in the DOS around 0.75 eV for \% strain above 0.25\% are reproduced in the predicted DOS curve. Fig. \ref{fig:9}(d-f) demonstrates strong correlations between DFT-calculated and RNN-predicted band edge and bandgap energies, validating the trained network and overall methodology.

\subsection{Dielectric constant calculation}
\begin{figure}[!ht]
    \centering
    \includegraphics[width=0.4\linewidth]{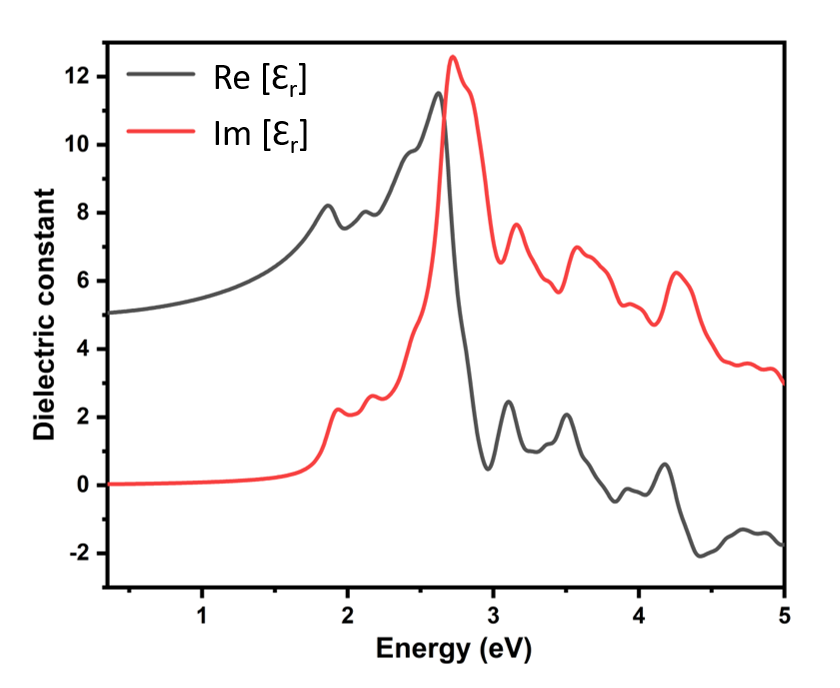}
    \caption{The real and imaginary part of the dielectric constant of pristine MoS$_2$.}
    \label{fig:Diel_p}
\end{figure}
Bending of the atomically thin lattice creates local distortions that reduces symmetry, which affects the dielectric constant of the system. The dielectric constant of the lattice affects electric screening of the system and might effect a change in the mobility of the system, via screening of charged scatterers. For calculating the dielectric constant, a uniform dense $k$ mesh has been used where the spin-orbit coupling has been considered. An intersmear value of $0.15ev$ was used, which is a broadening parameter for the interband contribution. 
Fig. \ref{fig:Diel_p} shows the real and imaginary part of dielectric constant of MoS$_2$ which is strongly consistent with previously reported results \cite{ng2022improving, Wang2022-di}. 
The atomically thin MoS$_2$ monolayer being a $2D$ system, is strongly anisotropic in its dielectric properties. Here, only the in-plane component of the dielectric constant has been taken into consideration. 

\subsection{Bending induced charge localization}
\begin{figure}[h!]
    \centering
    \includegraphics[width=\linewidth]{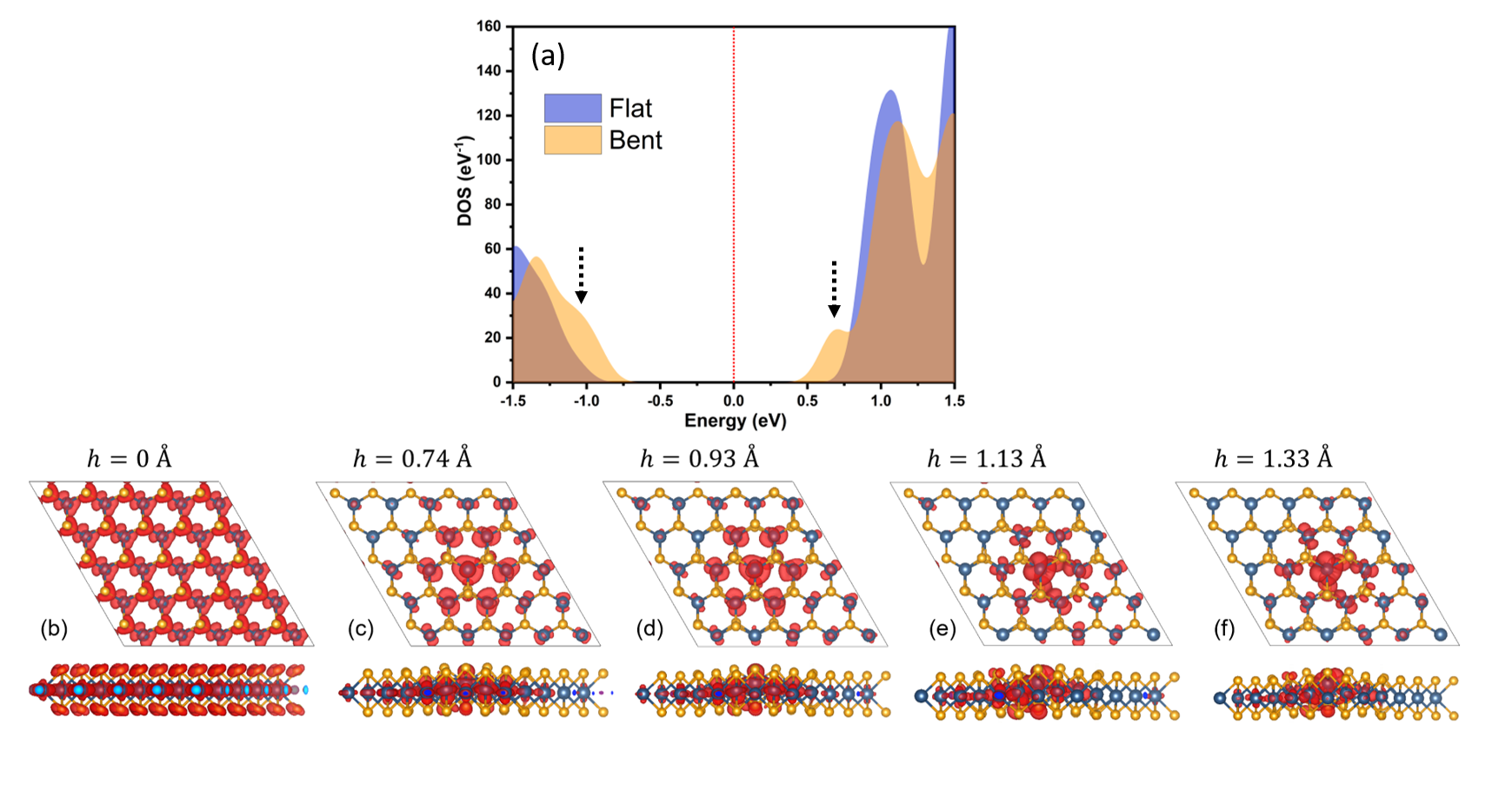}
    \caption{(a) Electronic DOS of flat and bi-axially bent MoS$_2$ with $h=1.33$ \AA, dotted arrows show the appearance of states at the band edge, (b-f) evolution of the partial charge density $|\Psi_{nk}|^2$ iso-surface corresponding to the band CBM edge state at the $\Gamma$ point with increasing deformation height.}
    \label{fig:CL}
\end{figure}
Due to this bending of the ML lattice and the ensuing modifications to the electronic structure, local charge density becomes inhomogeneous with higher density in regions of highest strain. Fig. \ref{fig:CL} shows that this inhomogeneity in charge density distribution increases with the increase in the height (i.e. strain) of the deformation. 
\section{Experimental Methods}
\subsection{Substrate patterning}
A $Si$ wafer was patterned with 500 nm diameter discs, of periodicity $2.5\mu m$ using standard e-beam lithography (Raith Pioneer 2) with PMMA resist. Post patterning, Cr and Au of thickness 5 nm and 45 nm were deposited using a thermal deposition system, followed by standard lift-off using acetone. 
\begin{figure}[!ht]
    \centering
    \includegraphics[width=0.75\linewidth]{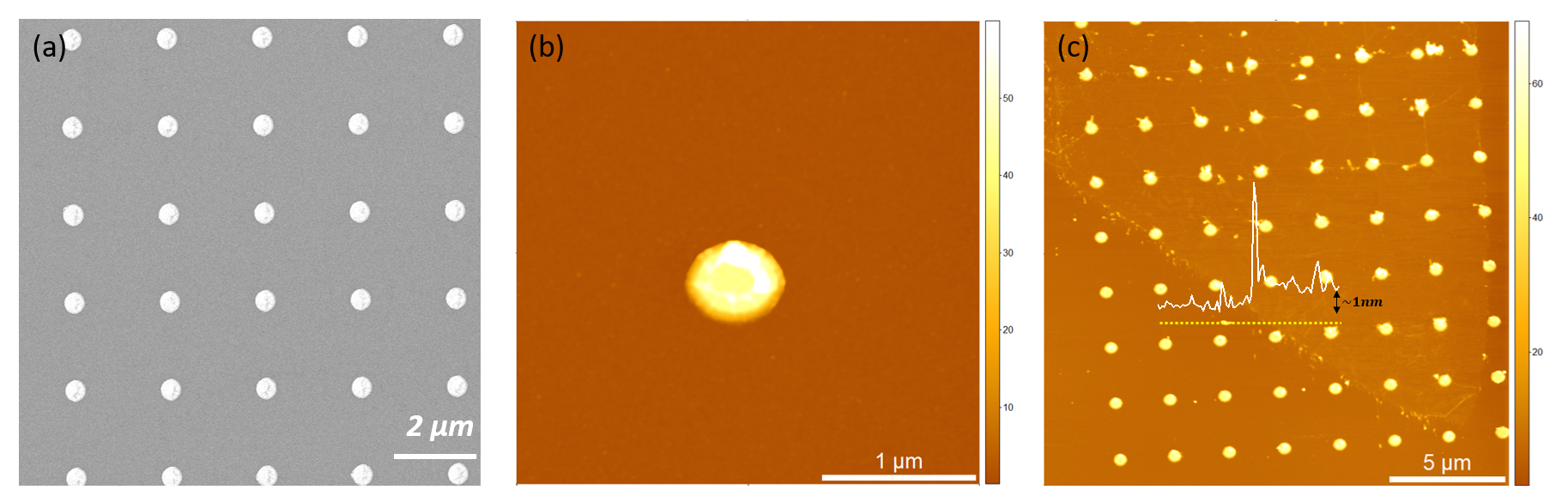}
    \caption{(a) Scanning electron microscopy image of the patterned substrate, (b) AFM image on a single pillar, (c) AFM topography after transferring a MoS$_2$ ML on top of the nano-pillar array. Inset shows a height profile line scan across the edge of the ML.}
    \label{fig:5}
\end{figure}
The SEM image (figure \ref{fig:5}a) confirms the periodicity of the nanostructures. The AFM topography of figure \ref{fig:5}b shows a top view of a pillar. The AFM topography of Figure \ref{fig:5}c evidences local deformations, wrinkles and bubbles originating from the pillars, and the line scan quantifies the thickness of the layer. 

\subsection{\texorpdfstring{MoS$_{2}$}{MoS2} Growth and Transferring}
The growth of the samples and their transfer have been carried out as previously reported following Kayal et. al\cite{kayal2023mobility}. The \(MoS_2\) flakes were grown by chemical vapor deposition (CVD) technique, where \(MoO_3\) and \(S\) powder was used as precursors. First, the \(MoO_3\) powder has been dissolved in ethanol and spin-coated on \(SiO_2/Si\) substrate. This substrate has been kept for heating in a quartz tube furnace. 500 mg of $S$-powder was kept at the upstream end of the tube. Argon flow of 100 $sccm$ was set and under $850^0 C$ of temperature, \(MoS_2\) growing happened. 

The as-grown \(MoS_2\) flakes were transferred onto the patterned substrate by wet etching treatment. First, a solution of 120K PMMA dissolved in 10 $wt\%$ anisole was prepared. The \(MoS_2\) grown $SiO_2/Si$ substrate was spin-coated with the solution and baked at $140^0 C$ for 15 minutes. The coated substrates were kept on $2M$ aqueous $NaOH$ solution at $60^0 C$, to etch out the $SiO_2$ layer and detach the fakes from the substrate into the PMMA layer. The detached film was scooped out of the solution and transferred to a water bath, and this job was repeated 3 times, and during each of the cycles, the film was kept for 30 minutes in the water. After that, the film was transferred onto the periodically patterned $Si$ substrate by scooping it from the water bath and subsequently baked at $80^0 C$ for 2 hours to evaporate the water. Finally, the flakes were cleaned from PMMA by dipping the substrate in warm acetone, followed by IPA cleaning and air blowing.
\begin{figure}[h!]
    \centering
    \includegraphics[width=0.55\linewidth]{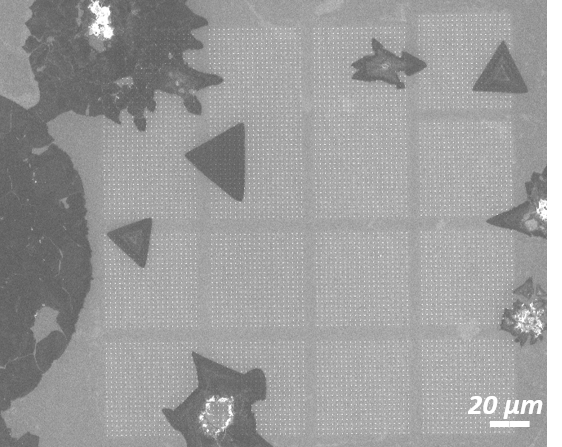}
    \caption{SEM image after transferring of the \(MoS_2\) flakes on the nano-structure}
    \label{fig:11}
\end{figure}

\subsection{Details of Atomic Force Microscopy (AFM) measurement}
All of the AFM measurements were performed using Bruker Multimode 8 AFM in ambient conditions. Morphological characterisation was conducted in the tapping mode, using non-coated cantilever with $f_0 \sim 325$ kHz from MikroMasch (HQ:NSC15/Al-BS). The conducting AFM current map was recorded using PeakForce TUNA  pre-amplifier, having variable voltage gain ($10^7 - 10^{10}$  V/A) with the capability of detecting current in the range: $100 fA - 1 \mu A$. The current map was recorded using Pt/Ir coated PFTUNA probes (k = 0.4 N/m) from Bruker. For the C-AFM maps, +0.2 V bias was applied to the sample and the probe was kept at virtual ground potential. 

\subsection{Photoluminescence and Raman Study}
A HORIBA Xplora Plus Raman setup based on a confocal microscope has been used for the spatially resolved photoluminescence and Raman study. The PL and Raman mapping was done with $532 nm$ laser excited through a $100x$ objective. The spot size of each of the data points was $0.5\mu m$. The grating which has been used for the PL map has a ruling density of $600gr/mm$, and for Raman has a ruling density of $2400gr/mm$.
\begin{figure}[!ht]
    \centering
    \includegraphics[width=\linewidth]{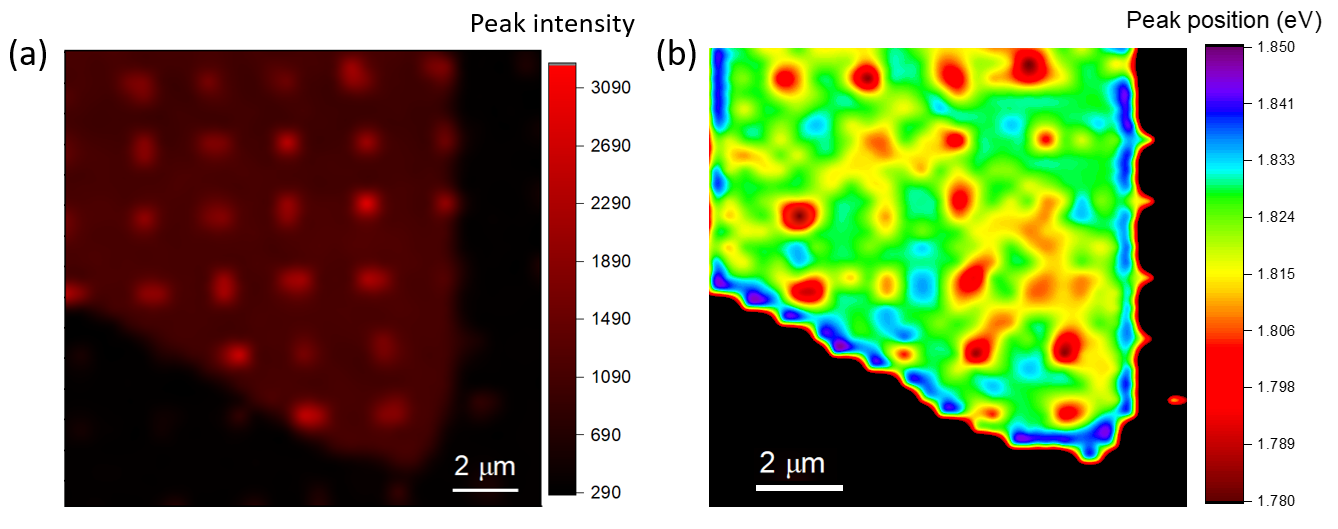}
    \caption{Spatial maps of (a) PL intensity; (b) PL peak energy}
    \label{fig:10}
\end{figure}
Figure \ref{fig:10} shows the PL peak intensity and peak position maps. The intensity increases on top of the pillars because of charge funneling at the bent region and the increase of the emission centers. 

\subsection{Strain calculation from Raman study}
A monolayer \(MoS_2\) has two distinguishable characteristic peaks, those are \(E_2g^1\) (in-plane vibration mode) and \(A_1g\) (out-of-plane vibration mode), and both of these are sensitive to strain, but the rate of shift is different for the peaks. 
\begin{figure}[!ht]
    \centering
    \includegraphics[width=\linewidth]{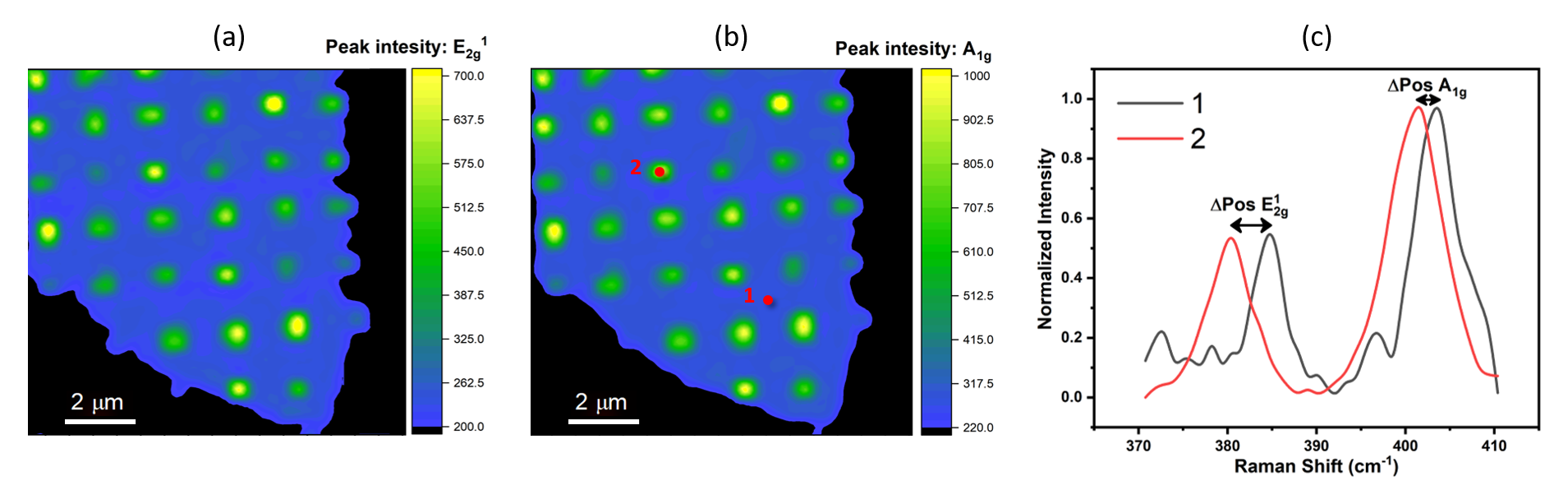}
    \caption{Spatial intensity map of (a) $E_{2g}^{1}$ mode and (b) $A_{1g}$ mode for ML MoS$_2$ on a patterned substrate (c) Raman spectra recorded on a strained and unstrained points, 1 and 2 shown in (b).}
    \label{fig:raman_intensity}
\end{figure}
A previously reported formula \cite{Zhang2024-bd} has been used here to calculate the \% of strain value from the relative shift of the Raman peak positions. The equation is given below,
\begin{equation}
\begin{pmatrix}
\Delta \text{Pos}\, E_{2g}^{1} \\
\Delta \text{Pos}\, A_{1g}
\end{pmatrix}
=
\begin{pmatrix}
-2\gamma_{E_{2g}^{1}} \,\text{Pos}E_{2g}^{1} & k_{n,E_{2g}^{1}} \\
-2\gamma_{A_{1g}} \,\text{Pos}A_{1g} & k_{n,A_{1g}}
\end{pmatrix}
\begin{pmatrix}
\varepsilon \\
n
\end{pmatrix}
\label{eqnS1}
\end{equation}
\noindent
Where $\varepsilon$ is the \% of strain, $n$ is the doping concentration, $\gamma$ is the Gr\"uneisen parameter (signifies the effect of strain on each of the vibration modes), and $k$ is the charge doping shift coefficients for each of the modes. For monolayer of \(MoS_2\), the parameters have been taken from some previous studies \cite{Zhang2024-bd, Michail2016-qu, Lloyd2016-hz}, the values are given below,
\[
\gamma_{E_{2g}^{1}} = 0.68, \gamma_{A_{1g}} = 0.21,  \\
k_{n,E_{2g}^{1}} = 0.33/10^{13} cm, k_{n,A_{1g}} = 2.22/10^{13} cm
\]
Using these values and the equation, and based on the relative shift of the Raman peak on top of the pillars with respect to the flat region, the \% of strain map has been generated, which has later been compared with the low-resolution \% of strain map, measured by AFM.
\subsection{DOS prediction on Raman-assisted strain map}
\begin{figure}[h!]
    \centering
    \includegraphics[width=\linewidth]{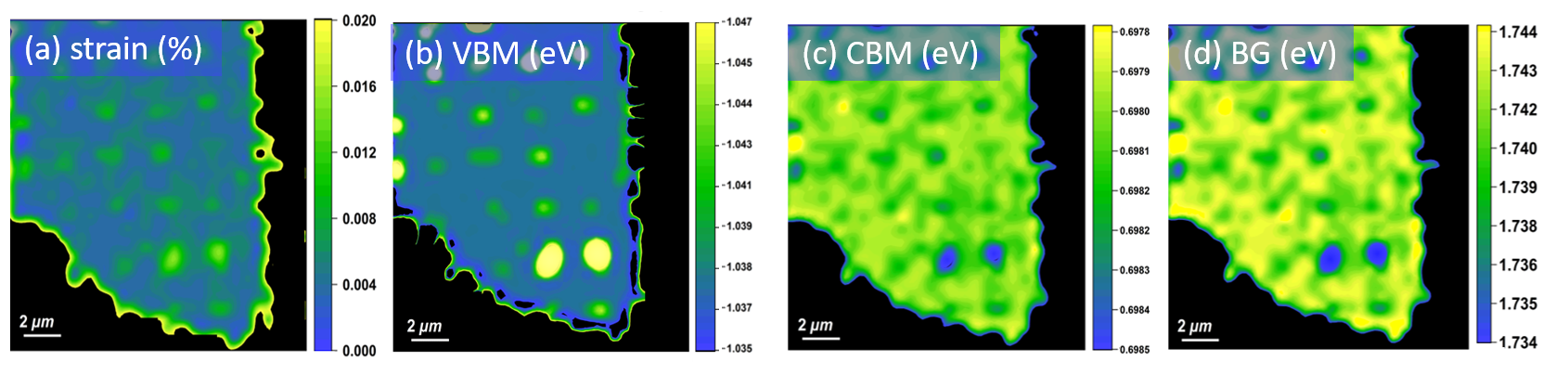}
    \caption{(a) Spatial maps of \% of strain value of ML MoS$_2$ on the pattern substrate (calculated based on relative raman peak shifting with respected of flat ML MoS$_2$) (b, c) spatial variation of Valance Band (VB) and Conduction Band (CB) edge position, collected from the predicted DOS map (d) spatial bandgap map}
    \label{fig:raman_strain}
\end{figure}
Using the equation \ref{eqnS1} of SI, the spatial strain map (fig. \ref{fig:raman_strain}(a) was generated using spatial Raman spectra. This strain map was further used to predict local DOS using the trained RNN. Fig. \ref{fig:raman_strain}(b and c) shows the spatial map of valance band (VB) and conduction band (CB) edge, and fig. \ref{fig:raman_strain}(d) shows the bandgap map, which is further compared with the PL peak position map (fig. 6).

\begin{figure}[h!]
    \centering
    \includegraphics[width=0.8\linewidth]{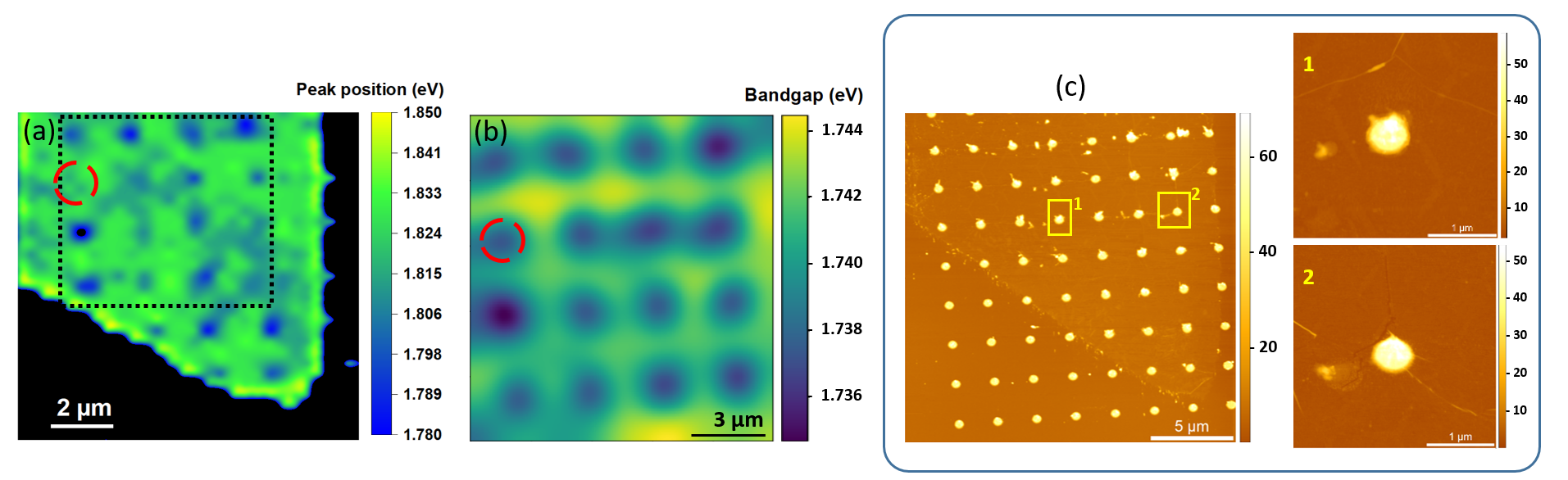}
    \caption{(a) Spatial maps of PL peak position energy of the ML MoS$_2$ flake on patterned substrate, (b) Low resolution band gap map, predicted by RNN (c) AFM topography with the zoomed views (1: The pillar, which does not show any red shift, 2: an example of an average kind of pillar, showing red shift).}
    \label{fig:12}
\end{figure}

The prediction map has been generated based on the strain map, which has been calculated from the AFM topography. The region in the AFM image where the MoS$_2$ flake is not present also have some topography change that will show some strain value, which will reflect on the bandgap map, and that bandgap will not be a correct one because of the absence of the flake. To avoid this circumstance, the region has been selected only where the \(MoS_2\) flake is present. In the PL peak position map of Figure \ref{fig:12}, there is a pillar that does not show any red shift, whereas in the bandgap map, the reduced bandgap is visible on that pillar. This is happening due to the perforation of the flake on top of that pillar, which is confirmed from the zoomed AFM topography scan. There are some topography changes in the AFM image, so a reduced band gap appears in the prediction map at that position. 
\end{spacing}

\clearpage
\bibliography{bibfile}
\end{spacing}

\end{document}